\documentstyle[12pt,aaspp4]{article}
\begin{document}
\title{Environments of Redshift Survey Compact Groups of Galaxies}
\author{Elizabeth J. Barton}
\affil{Harvard-Smithsonian Center for Astrophysics}
\authoraddr{ebarton@cfa.harvard.edu, 60 Garden St., Cambridge, MA 02138}
\and
\author{Reinaldo R. de Carvalho}
\affil{Observat\'orio Nacional / DAF}
\authoraddr{reinaldo@maxwell.on.br, Rua Gal. Jos\'e Cristino 77, 20921-400, Rio de Janeiro - RJ Brazil}
\and
\author{Margaret J. Geller}
\affil{Harvard-Smithsonian Center for Astrophysics}
\authoraddr{mgeller@cfa.harvard.edu, 60 Garden St., Cambridge, MA 02138}

\begin{abstract}

Redshift Survey Compact Groups (RSCGs) are tight knots of $N \geq 3$ 
galaxies selected from the CfA2+SSRS2 redshift survey.
The selection is based on physical extent and
association in redshift space alone.
We measured 300 new redshifts of fainter galaxies within 
$1 {\rm h}^{-1}{\rm\ Mpc}$ of 14 RSCGs to explore the relationship 
between RSCGs and their environments.
 
13 of 14 RSCGs are embedded in overdense regions of redshift space.  
The systems range from a loose group of 5 members to an Abell cluster. 
The remaining group, RSCG~64, appears isolated.
 
RSCGs are isolated and distinct from their surroundings 
to varying degrees, as are the Hickson Compact Groups.  
Among the 13 embedded RSCGs, 3 are distinct from their 
general environments (RSCG~9, RSCG~11 and RSCG~85).  

\end{abstract}


\section{Introduction}

Compact groups, the densest known systems of galaxies in the universe, are
apparent knots on the sky where member galaxies may be close enough to
interact and merge.  Compact groups were originally selected as apparently
dense systems on the sky (Rose 1977; Hickson 1982; Prandoni {\it et al.}
1994; see Hickson (1997) for a review).  More recently, Barton {\it et
al.} (1996) identified an objectively-selected sample of Redshift Survey
Compact Groups (RSCGs) from the CfA2 and SSRS2 magnitude-limited redshift
surveys.  The physical properties of RSCGs (velocity dispersion, density,
membership distribution) are similar to those of the Hickson Compact
Groups (HCGs). RSCG selection criteria include only physical extent and
association in redshift space. 

The abundance of compact groups is a challenge for dynamical models
because their crossing times are often much less than the Hubble time.
Some simulations and observations suggest that actual galaxy merging times
are longer. Simulated compact groups may take up to a few Gyr to merge if
a substantial amount of the group mass is located in the common group
potential well (Mamon 1987, Barnes 1989, Bode {\it et al.} 1993), or
possibly longer if the galaxies have a range of masses on particular
quasi-stable orbits (Governato {\it et al.} 1991). Pildis (1995) reports
evidence that the diffuse light in HCG 94 traces the same group potential
as the hot gas, suggesting a stable group potential on timescales $\geq$ 1
Gyr. 

Short lifetimes are not a problem if the groups form continually in dense
environments like loose groups (Barnes 1989;  Diaferio {\it et al.} 1994),
or if they are chance projections of galaxies and thus less dense than
they appear on the sky. Mamon (1986) suggested that about half of compact
groups are chance alignments of galaxies within loose groups, not physical
subcondensations. Similarly, Hernquist {\it et al.} (1995) proposed that
some compact groups are superpositions of galaxies viewed along filaments.
In these three scenarios, compact groups are embedded in environments that
are overdense in redshift space.  If some are collapsing physical systems
forming in loose groups, they will on average bear a different
relationship to their environments than if they are chance projections.
The environments of compact groups thus provide clues about the likelihood
that they are physically dense. 

Previous studies of galaxies near HCGs led to mixed conclusions about
their surroundings.  Sulentic (1987), Rood \& Williams (1989), de Carvalho
{\it et al.} (1994) and Palumbo {\it et al.} (1995) examined the
distribution of galaxies on the sky around HCGs; Rubin {\it et al.}
(1991), Ramella {\it et al.} (1994) and de Carvalho {\it et al.} (1997)
examined HCG environments in redshift space. These studies generally
conclude that some fraction of HCGs are embedded in denser environments,
with varying isolation from their environments.  These results raise the
questions (1) what does ``isolation'' mean for a compact group and (2) how
does Hickson's isolation criterion affect his sample?  The RSCG catalog
provides an approach to this issue; in contrast with Hickson, Barton {\it
et al.} included no isolation criteria in their sample selection. 

Catalogs of compact groups contain a mixture of systems.  When we refer to
a ``compact group'' we refer to a member of a catalog, a member which may
differ from all the others in fundamental ways and which may or may not be
a physically associated system.  Compact group environment studies seek to
answer two distinct questions:  (1) what are the environments of compact
group catalog members and (2) are individual compact groups distinct,
gravitationally bound systems or subsystems? In almost all cases we cannot
answer the latter question definitively without precise distance
measurements. 

Tidal distortions of member galaxies and x-ray emission are indicators of
a gravitationally bound system.  However, tidal distortions are not a
necessary consequence of a gravitationally bound system and x-ray emission
may be associated with individual galaxies in an unbound system. Nor do
luminosity function or morphological distinctions between ``field'' and
compact group galaxies indicate that individual compact groups are bound.
They can show only that the set of compact group galaxies differs from the
typical population. 

Optical galaxy distribution studies provide a statistical measure of the
probability that compact groups are physical systems. Here we characterize
the environments of 14 RSCGs with $cz > 2300 {\rm\ km \ s}^{-1}$ from the
CfA2North and CfA2South redshift surveys in order to:  (1) characterize
the environments of RSCGs and (2) explore how distinct RSCGs are from
their environments, as a clue to whether they are chance projections. We
address the first issue by testing whether the environments are overdense
in redshift space.  We apply statistical measures of the relationship
between each compact group and its redshift space environment to explore
the second issue. 

In Sec. 2 we describe the subsample of RSCGs and the construction of
redshift catalogs around RSCGs.  Sec. 3 is a description of our method of
defining the RSCG environment.  
In Sec. 4 we
address the embeddings of RSCGs.  
Sec. 5 contains our evaluation of
individual RSCG embeddings;  this section addresses the distinction
between individual compact groups and their environments.  
We conclude in Sec. 6. 

\section{Selection and Construction of RSCG Environment Catalogs}

\begin{figure}
\centerline{\epsfysize=6in%
\epsffile{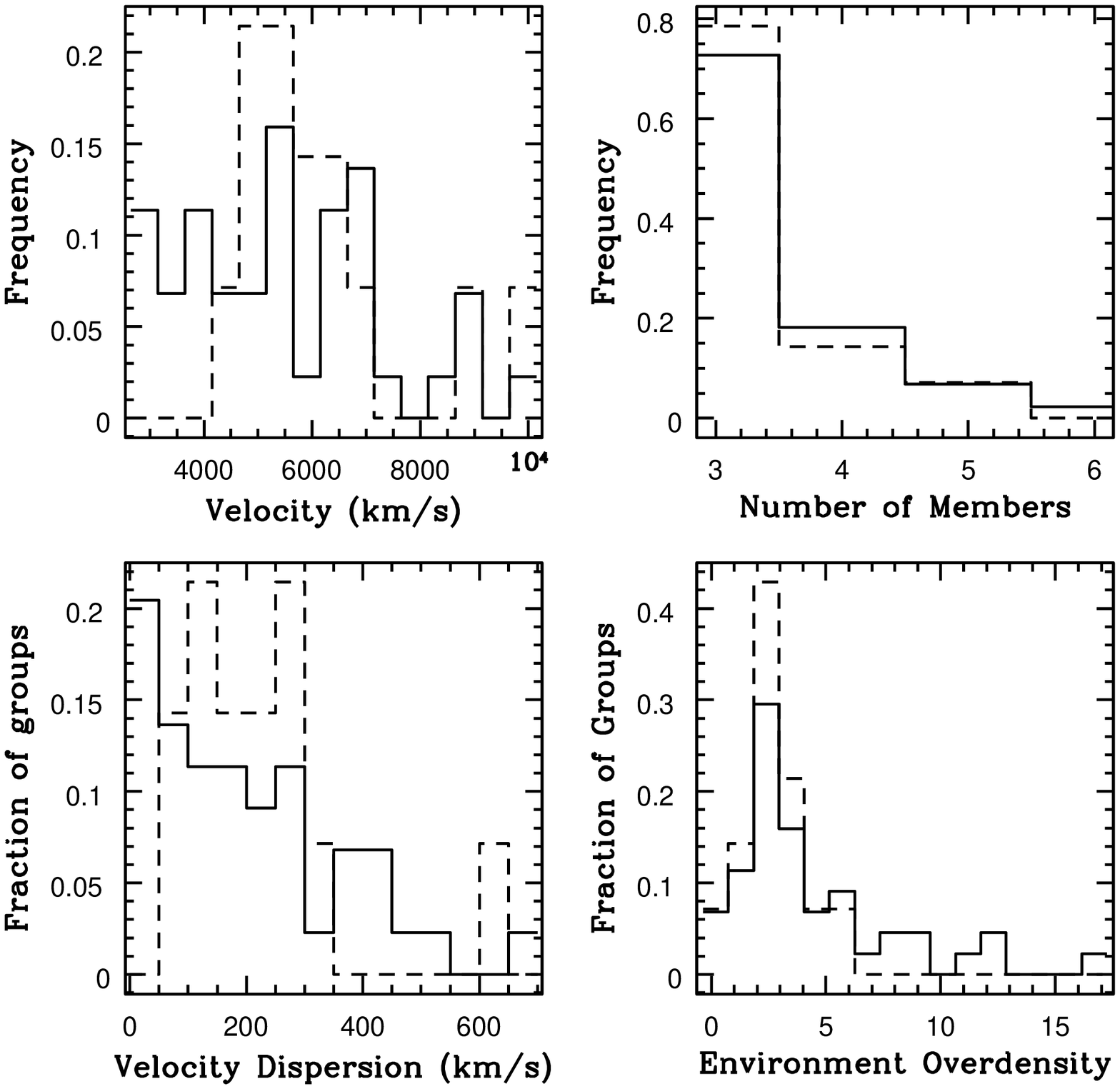}}
\caption{Distributions of various parameters, including the 44
 RSCGs in CfA2+SSRS2 with $cz > 2300 {\rm\ km\ s}^{-1}$ not 
included in this study (solid line) and the 14 RSCGs observed
here (dashed line): (a) redshift or velocity, (b) RSCG population,
(c) velocity dispersion and, (d) environment overdensity, 
$\frac{\langle \rho_{\rm env} \rangle}{\langle \overline{\rho} \rangle}$, 
as calculated in Barton {\it et al.} (1996).} 
\label{fig:compare}
\end{figure}

We select our subsample of 14 groups from the 47 RSCGs in the CfA2
redshift survey with $cz > 2300 {\rm\ km\ s}^{-1}$. We choose groups
located on POSS-II plates for which object catalogs are available (except
RSCG~29). The 14 RSCGs are marginally representative of the larger sample
of all 58 RSCGs in the CfA2+SSRS2 survey with $cz > 2300 {\rm\ km\
s}^{-1}$.  Table~\ref{tab:ks} lists the K-S probabilities that several
RSCG parameters have similar distributions in the observed subsample and
the sample of 44 RSCGs with $cz > 2300 {\rm\ km\ s}^{-1}$ in the
CfA2+SSRS2 survey. Figure~\ref{fig:compare} compares the distributions of
velocity (redshift), membership frequency, velocity dispersion and
overdensity of the environment (compared with the average over the
redshift survey) for the two subsamples.  Most of the 14 RSCGs are in
dense regions of the redshift survey.  Our sample excludes the densest
environments. 

We extract catalogs of all objects from the Digitized POSS-II sky survey
(Djorgovski {\it et al.} 1997) to a limiting magnitude of $m_{\rm lim}
\sim 16.9$ in g to include the faintest galaxies that we can observe
efficiently with the Tillinghast telescope.  We use SKICAT object
classifications, which are based on a Decision Tree algorithm (Weir {\it
et al.} 1995; Weir 1994), to identify a sample of 573 galaxies within $1
{\rm\ h^{-1}\ Mpc}$ (projected) of the center for 13 of the RSCGs in our
subsample.  Because we are looking for relatively bright objects, for
which SKICAT classifications are the most uncertain, we examined either
the POSS-I or POSS-II image of each object to check the SKICAT
classifications.  We also checked the regions for bright galaxies missed
by the SKICAT algorithm.  For the remaining group, RSCG~29, we used the
FOCAS object identification package in IRAF on the Digitized Sky Survey
image to identify 30 nearby galaxies. 

To avoid remeasuring known redshifts, we checked the CfA Redshift
Catalogue (Geller \& Huchra 1989; Huchra {\it et al.} 1990; Huchra {\it et
al.} 1995a; Huchra {\it et al.} 1995b;  Giovanelli \& Haynes 1985;
Giovanelli {\it et al.} 1986;  Haynes {\it et al.} 1988; Giovanelli \&
Haynes 1989;  Wegner {\it et al.} 1993; Giovanelli \& Haynes 1993; 
Vogeley 1993) for velocity measurements of the sample galaxies. In
ambiguous cases we remeasured velocities, including those for several RSCG
galaxies. Table~\ref{tab:obs} describes the measured sample, which
contains a total of 509 galaxies, including 300 newly measured galaxies. 
In order to save space and avoid redundant publication of data,
Table~\ref{tab:glx} lists only the newly measured galaxies, and identifies
the galaxies in each RSCG or its environment, according to the criteria
described below.  A complete list of the galaxies in our catalog,
including new redshifts and redshifts taken from the CfA Redshift
Catalogue, is available via anonymous ftp at: 
ftp://cfa0.harvard.edu/pub/barton.  The RSCG
coordinates in the table differ from those in the original RSCG paper
because we now have coordinates good to $\sim 1$ arcsecond for RSCG~29
(POSS-I) or $\sim 0.5$ arcseconds for the other regions (POSS-II). 

We measured the new redshifts with the FAST spectrograph at the 1.5m
Tillinghast reflector on Mt. Hopkins.  We used a grating with 300 lines/mm
to disperse the light into the wavelength range $4000-7500$ \AA;  typical
exposure times were 10 - 20 minutes.  We measured radial velocities using
the XCSAO program in IRAF (Kurtz {\it et al.} 1992).  The program
implements the cross-correlation technique of Tonry \& Davis (1979) on
data binned logarithmically in wavelength. Errors in velocity for
emission-line redshifts are dominated by fluctuations in the small number
of emission regions contributing to the measurement. To account for this
effect empirically we add $75 {\rm\ km\ s}^{-1}$ in quadrature to the
cross-correlation errors for emission line redshifts (Kurtz {\it et al.},
private communication). We did not change original CfA2 Redshift Survey
errors, so errors for emission redshifts may be underestimated.

\section{Identifying Systems Surrounding the RSCGs}

We implement a slight modification of the friends-of-friends group-finding
algorithm with ``volume scaling'' to identify members of loose systems
around the RSCGs (Huchra \& Geller 1982). We use a code from Ramella {\it
et al.} (1997).  We identify galaxy systems as linked sets of
``neighboring'' galaxies.  To determine whether two galaxies belong to the
same system, we consider both their projected separation, $\Delta D$, and
their line-of-sight velocity difference, $\Delta V$.  At low redshift,
$\Delta D = 2 \left(\frac{v}{H_0}\right) \sin(\frac{\Delta \theta}{2}),$
where $\Delta \theta$ is the angular separation on the sky and $v=cz$ is
the average redshift.  We scale $\Delta D$ and $\Delta V$ in accord with
the sampling of the luminosity function.  The volume we search for
``neighbors'' is inversely proportional to the integral of the luminosity
function at the median redshift of the RSCG. Throughout the paper we use
$H_0 = 100 {\rm\ km\ s}^{-1}{\rm\ Mpc}^{-1}$. 

We restrict the density contrast of our groups to $\delta \rho / \rho \geq
80$ by specifying fiducial parameters, $D_0$ and $V_0$, and requiring
$\Delta D \leq R D_0$ and $\Delta V \leq R V_0$, where $R$ is the
redshift-dependent scaling parameter.  $D_{0}$ and $R$ are functions of
the limiting Zwicky magnitude, $m_{\rm lim,Zw}$.  As we lack photometric
calibration for the object catalogs, our sample is inhomogeneous; $m_{\rm
lim, Zw}$, and therefore $D_{0}$ and $R$, vary among the RSCG
environments. Linked sets of ``neighbors'' satisfying these criteria are
part of the same system.  Here, $R$ depends on the median velocity of the
RSCG: 
\begin{equation} 
R = \left[ \int^{M_{\rm med}}_{-\infty} \Phi (M) dM/ \int^{M_{\rm lim}}_{-\infty} \Phi(M) dM\right]^{-1/3}, 
\end{equation}
where $M_{\rm med} = m_{\rm lim, Zw} - 25 - 5\log(\frac{v_{\rm
med}}{H_0})$ is the limiting absolute magnitude at the median group
velocity, $v_{\rm med}$; $\Phi(M)$ is the CfA2North or CfA2South
luminosity function (Marzke {\it et al.} 1994) and $\frac{v_{\rm
med}}{H_0}$ is in Mpc.  Similarly, $M_{\rm lim} = m_{\rm lim, Zw} - 25
-5\log(\frac{v_F}{H_0})$, where $v_F$ is an arbitrary fiducial velocity. 
We choose $v_F = 1000\ {\rm km\ s}^{-1}$. 

The parameter $D_0$ determines the minimum galaxy density enhancement,
$\delta \rho / \rho$, of systems we identify.  We use a different $D_0$
for each field, ranging from $\sim 220$--$360 {\rm\ kpc}$, corresponding
to $\delta \rho / \rho = 80$, in accord with Ramella {\it et al.} (1989). 
We adopt $V_0=350 {\rm\ km\ s}^{-1}$ to prevent groups from spanning voids
but to allow large velocity dispersion systems.  Barton {\it et al.}
(1996) used the friends-of-friends algorithm with $D_0 = 50 {\rm\ kpc}$
and $V_0 = 1000 {\rm\ km\ s}^{-1}$ and no volume-scaling ($R = 1$) to
identify the original RSCG sample from the CfA2+SSRS2 redshift survey.
Because we search a limited region on the sky, we may miss parts of the
galaxy systems that contain the RSCGs. 

We estimate the effective Zwicky limiting magnitude, $m_{\rm lim, Zw}$, of
each RSCG environment region.  Using only the galaxies for which we know
both SKICAT instrumental g magnitudes and Zwicky magnitudes, we estimate
the relationship between the two for each region separately using a linear
least-squares fit.  Because of confusion in the region of Abell 194, we
use only a restricted sample in the regions of RSCGs 10 and 11, based on
the catalog of Chapman {\it et al.} (1988). Table~\ref{tab:fofa} lists the
results for each region. 

We choose the magnitude of the faintest galaxy with a redshift as our
limiting magnitude.  The completeness of each region to the limiting
magnitude is listed in the last column of Table~\ref{tab:obs}.  For the
most incomplete regions, we test the effects of the choice of limiting
magnitude on the galaxy environments.  We find that it has no effect for
most regions, and no qualitative effects for any regions. 

The limiting projected separation we adopt for each environment is more
generous than the criterion applied to find the RSCGs and the velocity
separation criterion is more strict.  Table~\ref{tab:fofa} lists the
values of $R D_0$ and $R V_0$.  In all cases, the RSCG galaxies are
``neighbors''.  RSCG~64 is the only system where there are no other
galaxies in the environment.  Throughout the paper, we refer to the looser
aggregate of galaxies identified by the algorithm as the environment of
the RSCG. 

\section{Are Apparent Compact Groups Embedded in Dense Environments?}

Previous studies of compact group environments yield an inconsistent
picture of the embedding of compact groups.  These inconsistencies
originate from incomplete data sets along with the assumptions underlying
some analyses.  For example, some studies argue that a surrounding loose
group is not present, based on the distribution of surrounding galaxies on
the sky alone.  In fact, loose groups are often hard to distinguish from
the foreground/background without redshifts. 

Studies done in redshift space are cleaner.  However, the data must be
complete to evaluate the statistical significance of detection. Rubin {\it
et al.} (1991) examined the incomplete CfA Redshift Survey Catalogue
(Huchra {\it et al.} 1991) within 1000~km~s$^{-1}$ and 2.8~h$^{-1}$~Mpc of
21 HCGs with mixed results.  They could not evaluate the significance of
the general absence of surrounding loose groups because of the
incompleteness of the catalog. 

Ramella {\it et al.} (1994) extracted galaxies within 1.5~h$^{-1}$~Mpc and
1500~km~s$^{-1}$ of 38 HCGs from the CfA2 complete, magnitude-limited
redshift survey.  They compared the number of detected galaxies, $N_{\rm
n}$, to the number of galaxies expected in the region, $N_{\rm int}$.  29
HCGs have $N_{\rm int} \gg N_{\rm n}$.  The properties of the surrounding
systems are similar to those of loose groups extracted from the redshift
survey by Ramella {\it et al.} (1989). Barton {\it et al.} (1996)
extracted the same-sized regions around the RSCGs and obtained a similar
result: of the more distant RSCGs ($v \geq 2300 {\rm\ km\ s}^{-1}$), 72
\% (42/58) have $N_{\rm n} > 2N_{\rm int}$. 

Here, we again reach a similar conclusion: 13/14 RSCGs are embedded in
regions that would qualify as potentially bound systems according to
Ramella {\it et al.} ($\frac{\delta \rho}{\rho} \geq 80$ on the sky with
additional restrictions on velocity separation).  The richness and density
of these systems varies from loose groups of 5 members (RSCG~85) to an
Abell cluster (RSCG~10 and RSCG~11 in Abell~194).  The properties of these
systems undoubtedly vary.  Zabludoff \& Mulchaey (1997) and Mulchaey \&
Zabludoff (1997) use multi-fiber spectroscopy and ROSAT PSPC data to study
poor groups.  Some groups in their sample display properties similar to
x-ray clusters and others show no definitive evidence that they are bound. 

\section{Are RSCGs Distinct From Their Environments?}

We compare each RSCG with its surroundings to explore the probability that
an individual RSCG is a bound physical subsystem by asking whether
its redshift-space configuration is likely to arise by chance.  
We use two parameters, p($\Delta v_{\rm max}$) and $D_{\rm nn,s}$, as
partial diagnostics, in redshift and on the sky, respectively, of the
relationship between the RSCG and its environment.  
p($\Delta v_{\rm max}$) is a direct, but insensitive, measure
of the probability that the velocity distribution of the RSCG relative
to its environment arises by chance.  In contrast, $D_{\rm nn,s}$ only
ranks the groups according to their relative isolation from 
neighbors in their environments.

The function p($\Delta v_{\rm max}$) is
the probability that $N_{\rm cg}$ galaxies drawn from the observed 
velocity distribution of the environment have $\Delta v \leq 
\Delta v_{\rm max}$.  Here, $N_{\rm cg}$ is the number of galaxies in the
RSCG and $\Delta v_{\rm max}$ is the largest velocity difference between
members of the RSCG.  The environments of the RSCGs were chosen with 
stricter velocity separation criteria than the RSCGs 
($R V_0 \leq 1000 {\rm\ km\ s}^{-1}$ in Table~\ref{tab:fofa}). Therefore,
p($\Delta v_{\rm max}$) is an upper limit to the value it
would have if the environments and RSCGs were chosen with the same
velocity criteria. When small, p($\Delta v_{\rm max}$) is an
indicator of association within well-populated environments;  the probability
is then large that the RSCG is not just a chance superposition.  For
RSCGs in poor environments, the behavior of p($\Delta v_{\rm max}$) is
dominated by small number statistics and the statistic is not a good
discriminant. 

The parameter $D_{\rm nn}$ is the projected distance between the center of
the RSCG and the nearest neighbor within the environment.  A scaled
$D_{\rm nn, s} = D_{\rm nn} / R$ accounts for different absolute magnitude
limits within different systems.  Physically, this measure evaluates the
separation between the RSCG center and the nearest galaxy for an
equivalent group located at the fiducial velocity, $v_F = 1000 {\rm\ km\
s}^{-1}$.  This interpretation assumes a simple model for the galaxies in
the neighborhood --- the spatial distribution is random, luminosity and
position are uncorrelated and the luminosity function is the same around
every RSCG.  $D_{\rm nn,s}$ is useful only as an indicator of {\it relative}
compactness on the sky because it has not been calibrated on any 
complete model of loose groups.  

If RSCGs are collapsing subsystems embedded in looser
environments, they will on average be tighter on the sky than their
surrounding environments.  Any particular RSCG can have a high value
of $D_{\rm nn,s}$ by chance if it is only an apparent alignment, but
groups with high values of $D_{\rm nn,s}$ are less likely to be
alignments than other RSCGs.  The set of RSCGs with low values of 
$D_{\rm nn,s}$ may still contain physical subsystems --- they are merely
more likely to be contaminated with chance projections. 
We note that Hickson effectively chose only 
compact configurations with 
$D_{\rm nn} \geq 3 R_{\rm HCG}$ to minimize the number of chance
alignments, where $R_{\rm HCG}$ is the radius of the smallest
circle on the sky containing all of the HCG galaxy centers.  
Barton {\it et al.} (1996) argue that such a criterion may exclude 
real, physical systems located in dense environments.  They found such
an isolation criterion unnecessary because they selected the RSCGs based on
redshift separation and were therefore able to eliminate interlopers in
redshift space.  We compute $D_{\rm nn,s}$ for the RSCGs {\it a posteriori} 
to rank the groups as more or less likely accidental superpositions.

Table~\ref{tab:param} lists these statistics along with the number of
galaxies in the environment ($N_{\rm env}$) and the median velocity ($v_{\rm
med, env}$).  These
parameters refer only to the $1 {\rm\ h}^{-1}$~Mpc region we survey around
each RSCG. 

\begin{figure}
\plottwo{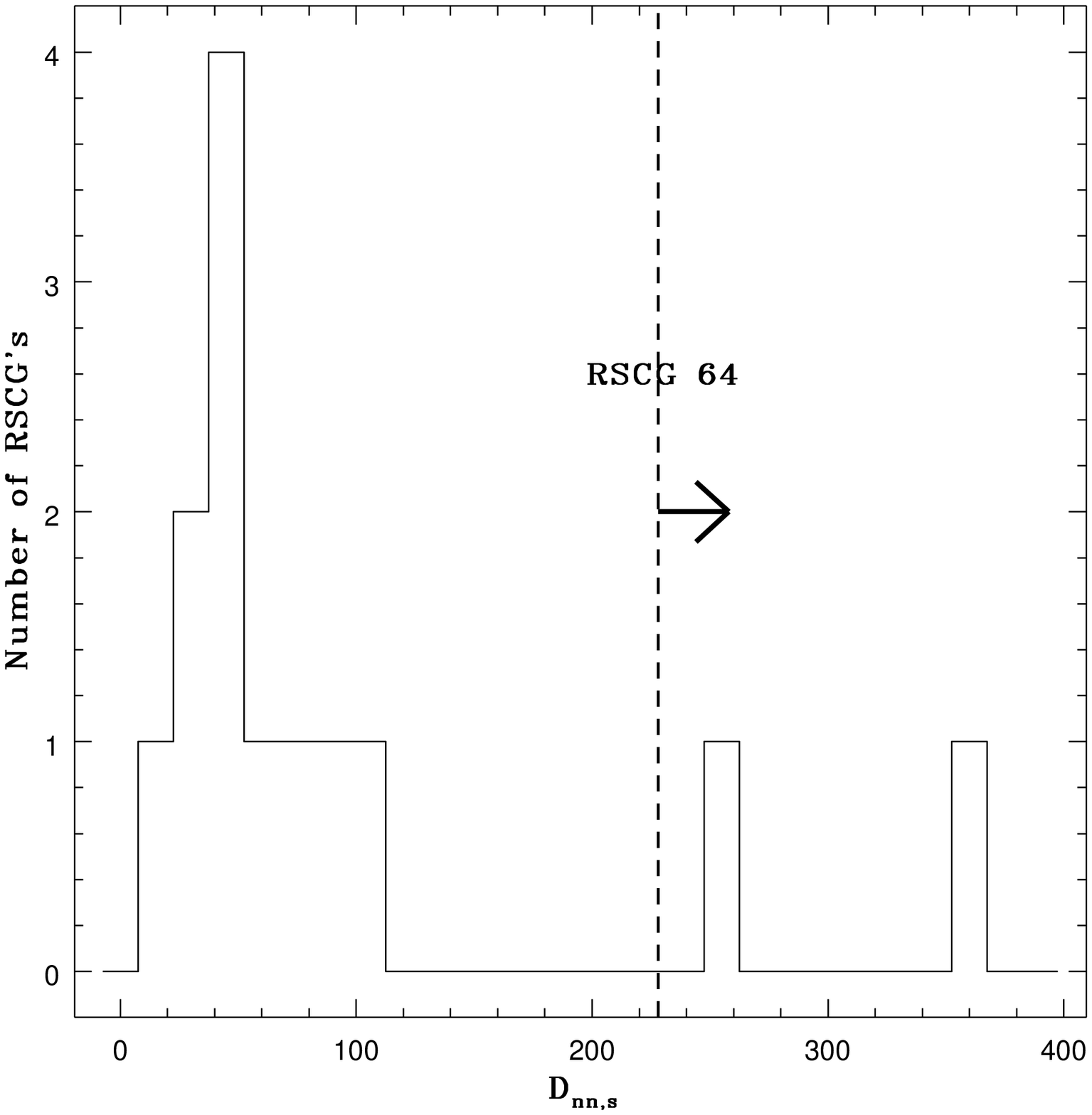}{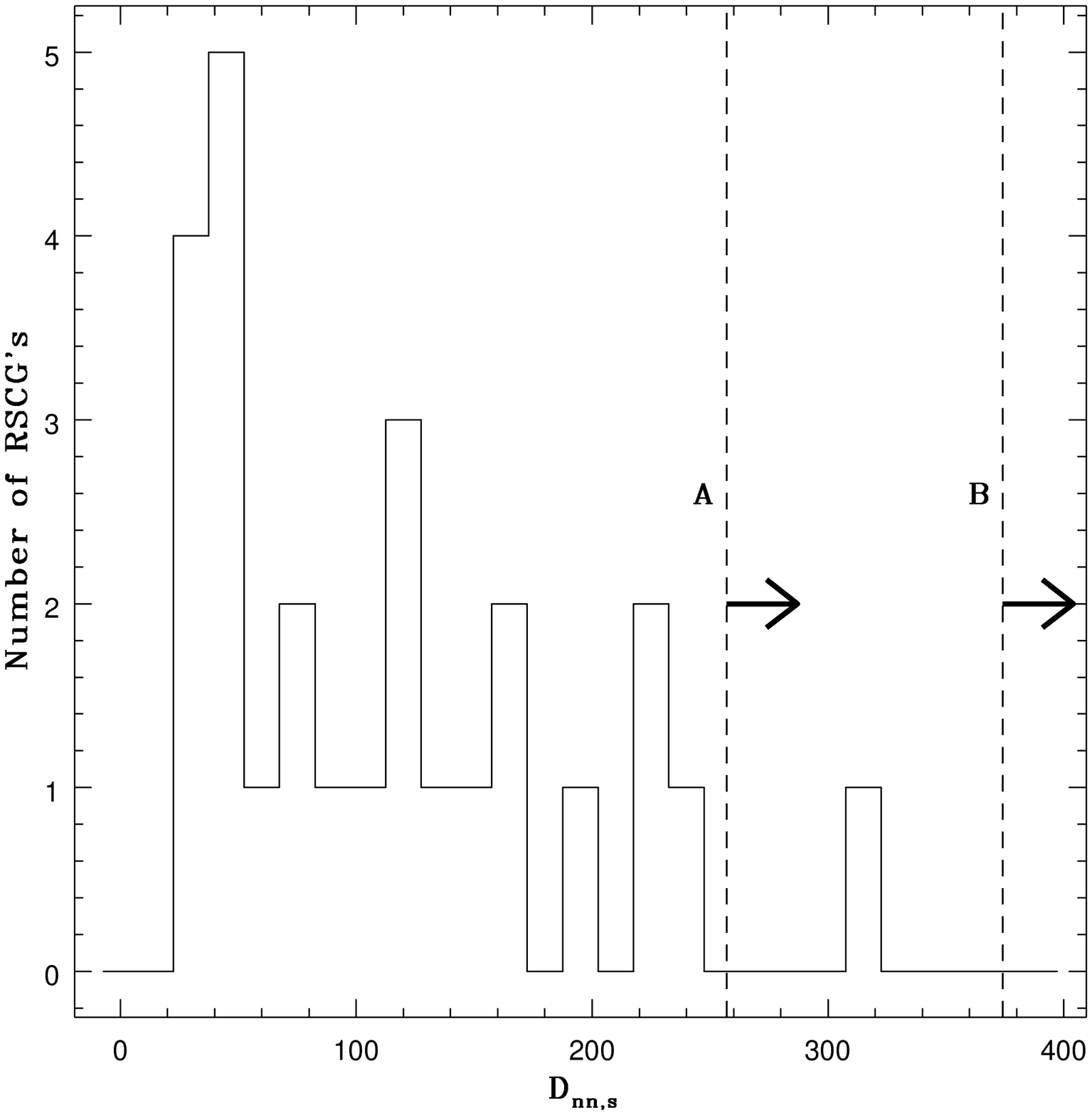}
\caption{$D_{\rm nn,s}$ distributions: (a) distribution of $D_{\rm nn,s}$ 
for the 14 RSCGs in our sample.
The outliers are RSCG~85 and  RSCG~9, 
with $D_{\rm nn,s} = 250$~scaled~kpc and $D_{\rm nn,s} = 359$~scaled~kpc,
respectively.  The vertical dashed line represents the 
lower limit of $D_{\rm nn,s}$ for
RSCG~64, and (b) distribution of $D_{\rm nn,s}$ for the 33
other RSCGs in the CfA2 survey with $cz \geq 2300 {\rm\ km\ s}^{-1}$.
26 have environment galaxies according to our criteria and
are included in the histogram; 7 have only upper limits, represented
by the dashed lines A (4 RSCGs in CfA2North) and B (3 RSCGs in CfA2South).}
\label{fig:dnns_hist}
\end{figure}

\begin{figure}
\centerline{\epsfysize=6in%
\epsffile{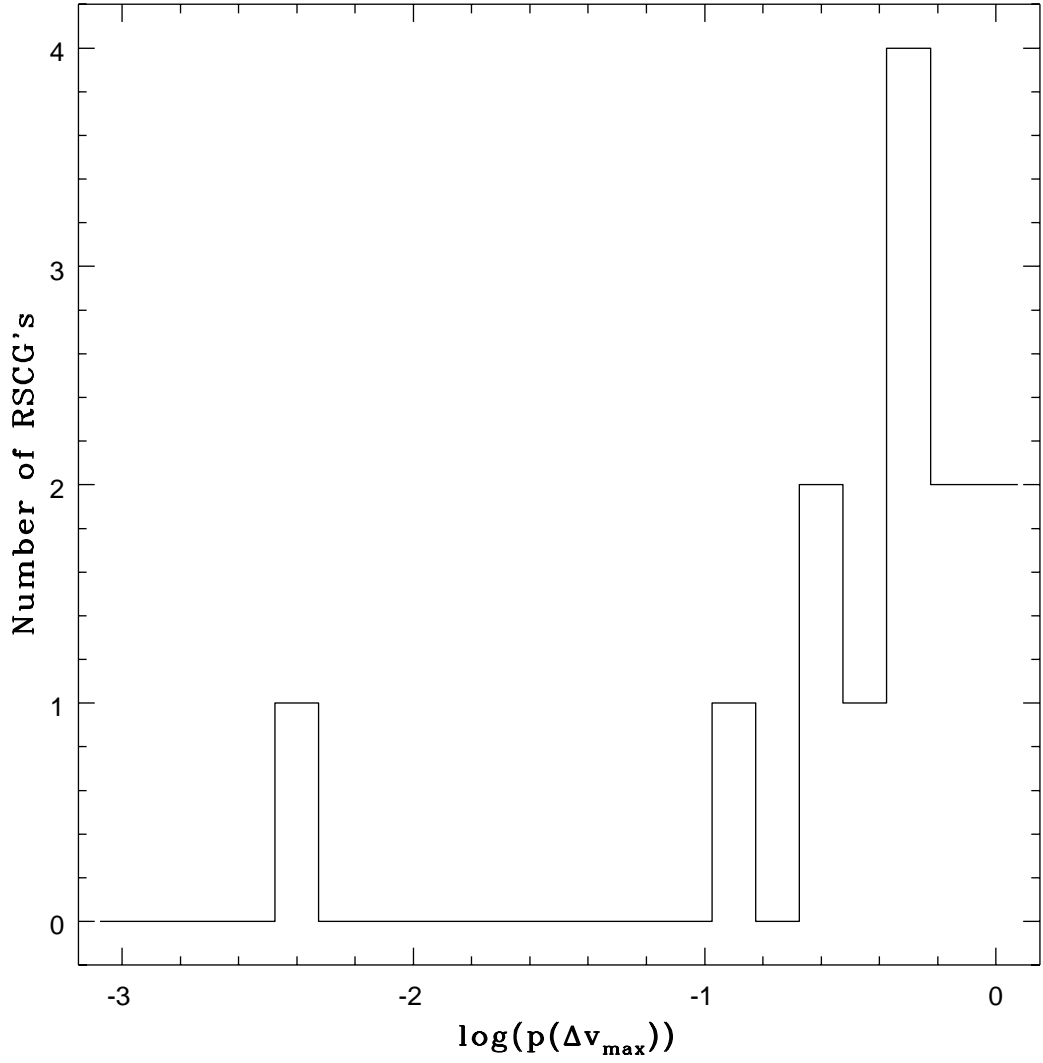}}
\caption{The distribution of $\log(p(\Delta v_{\rm max}))$ for 
the 14 RSCGs in our sample.
The outlier is RSCG~11, in Abell~194, with $p(\Delta v_{\rm max}) = 0.004$.}
\label{fig:pv_hist}
\end{figure}

Figs.~\ref{fig:dnns_hist}a and \ref{fig:pv_hist} show the sample
distributions of $D_{\rm nn,s}$ and log(p($\Delta v_{\rm max}$)),
respectively, for the embedded RSCGs.  Fig.~\ref{fig:dnns_hist}a shows
the lower limit to $D_{\rm nn,s}$ for the RSCG~64, which has an empty
neighborhood.  This limit is imposed by the friends-of-friends algorithm
and is equal to $D_0$.  The lower limit is well above the distribution of
$D_{\rm nn,s}$ for the majority of the sample. Fig.~\ref{fig:dnns_hist}b
shows the $D_{\rm nn,s}$ distribution of the remaining CfA2 RSCGs for
comparison, including lower limits for RSCGs with empty neighborhoods in
the CfA2 redshift survey.  Note that surrounding galaxies fainter than
$m_{\rm Zw} = 15.5$ are not included in Fig.~\ref{fig:dnns_hist}b. 

In the $D_{\rm nn,s}$ plot (Fig.~\ref{fig:dnns_hist}a), RSCG~9 and RSCG~85
are the outliers. They appear more isolated from the other
galaxies in their environments and thus less likely than the other 
RSCGs to be chance superpositions of galaxies within looser systems.  The 
distribution of $D_{\rm nn,s}$ for the whole RSCG catalog in 
Fig.~\ref{fig:dnns_hist}b is more spread out than the distribution for
the 14 RSCGs in this study.  This spreading may indicate that large
values of $D_{\rm nn,s}$ arise by chance.
In Fig.~\ref{fig:pv_hist}, RSCG~11, in
Abell~194, is the outlier; the galaxies have a very low probability 
($< 0.5 \%$) of being associated by chance.  RSCG~11 is surprisingly 
close to the center of the cluster both in velocity space and on the sky. 
It may be part of a cold core (e.g. Bothun \& Schombert 1988;  
Merrifield \& Kent 1991; Mohr
{\it et al.} 1996). The two large, elliptical galaxies in RSCG~11 appear
to be within a common envelope.  An additional large elliptical, with a
velocity equal to the median velocity of the group environment, lies
within 35.5~h$^{-1}$~kpc of a member of the RSCG. 

\begin{figure}
\centerline{\epsfysize=6in%
\epsffile{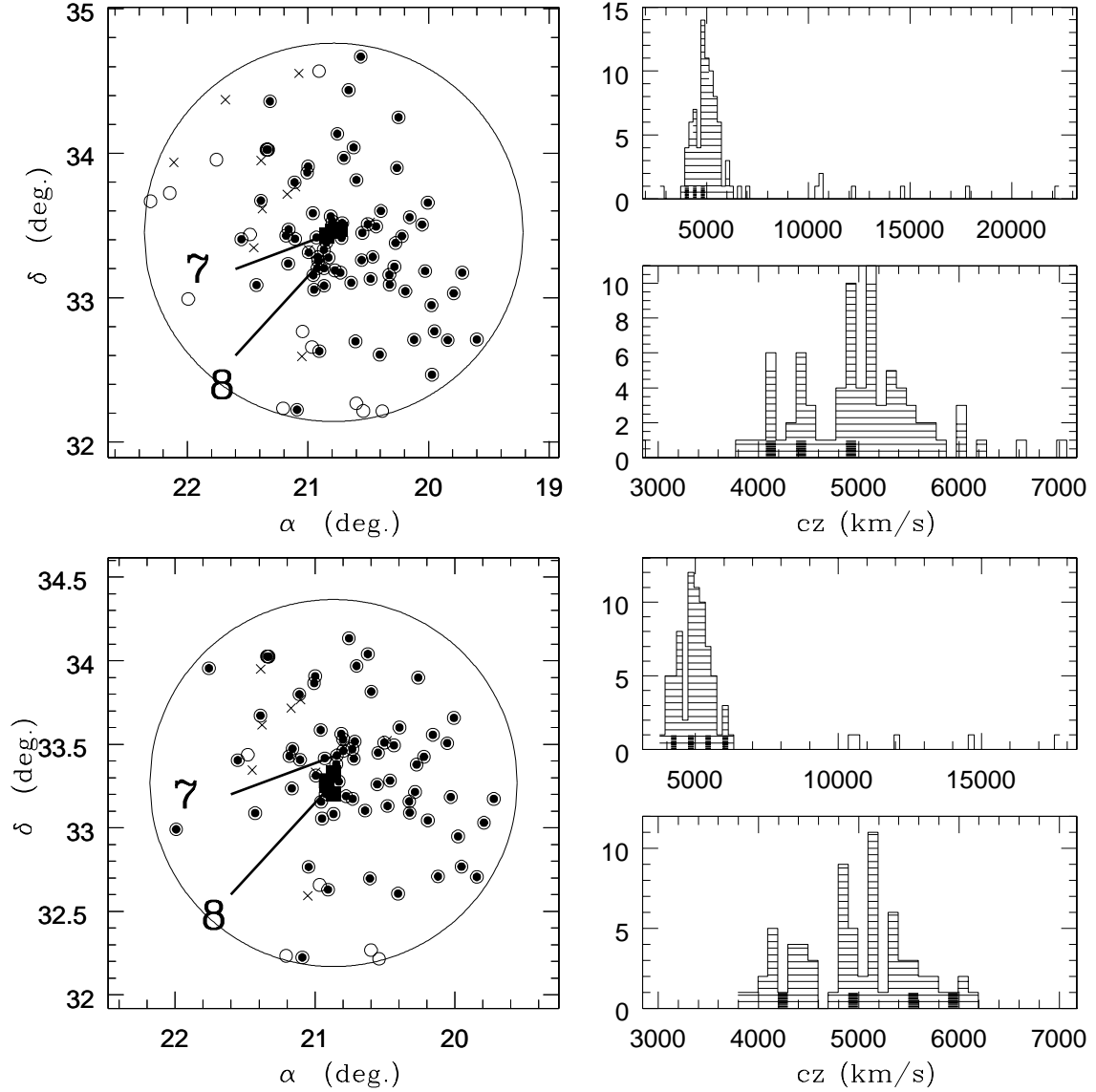}}
\caption{(a) RSCG~7 and (b) RSCG~8: (1) Galaxy positions on the sky (left).  
Filled squares are galaxies in the RSCG, filled circles are galaxies 
in the RSCG environment, empty circles are foreground/background 
galaxies and x's are galaxies without measured velocities; (2) 
velocity distributions (right).  The upper histogram includes
all galaxies with measured redshifts.  The lower histogram (right) expands 
the region around the RSCG. Lightly shaded regions are the
RSCG environment; heavily shaded regions are the RSCG itself.} 
\label{fig:845}
\end{figure}

\begin{figure}
\centerline{\epsfysize=6in%
\epsffile{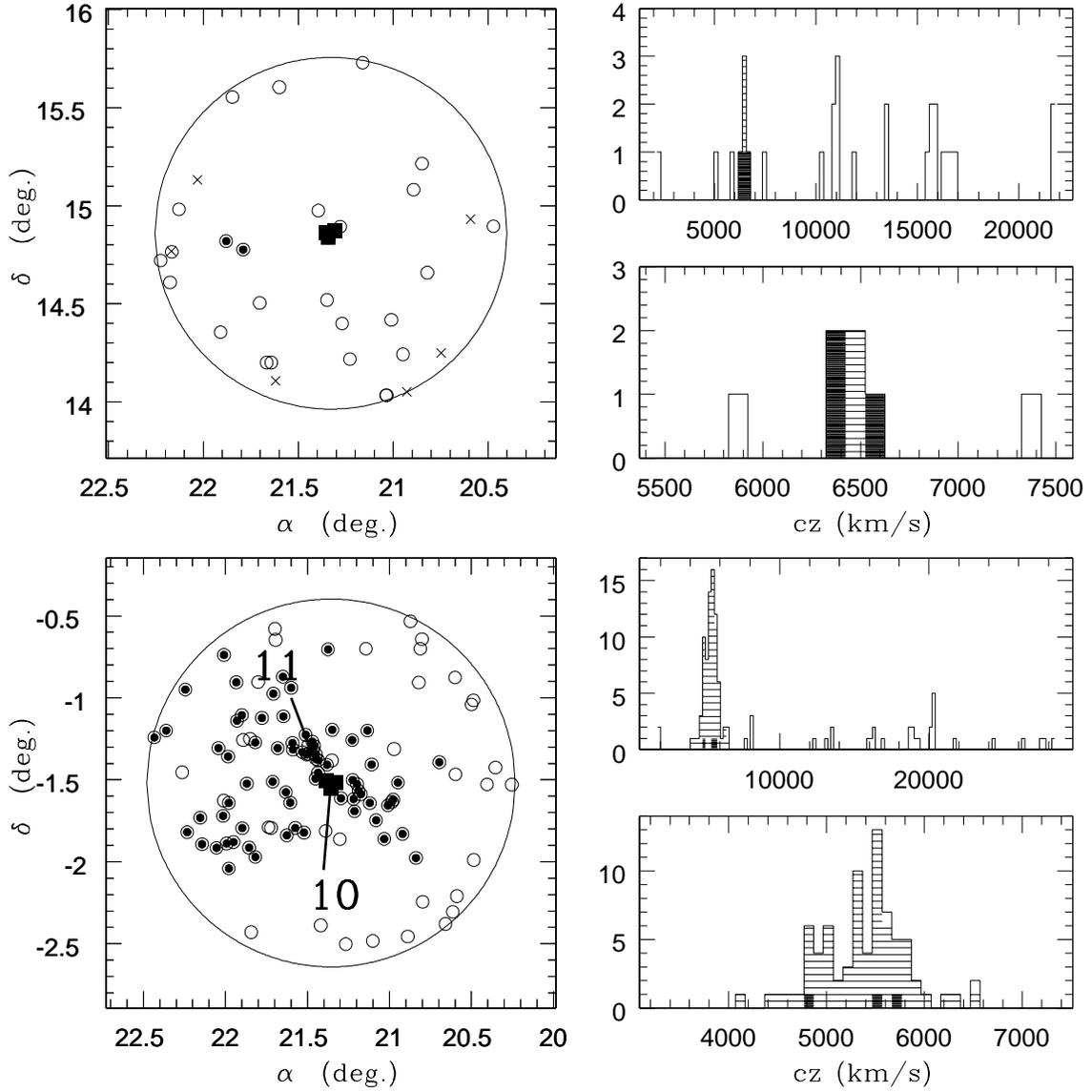}}
\caption{(a) RSCG 9 and (b) RSCG 10.  
Format as in Fig.~\ref{fig:845}.}
\label{fig:904}
\end{figure}

\begin{figure}
\centerline{\epsfysize=6in%
\epsffile{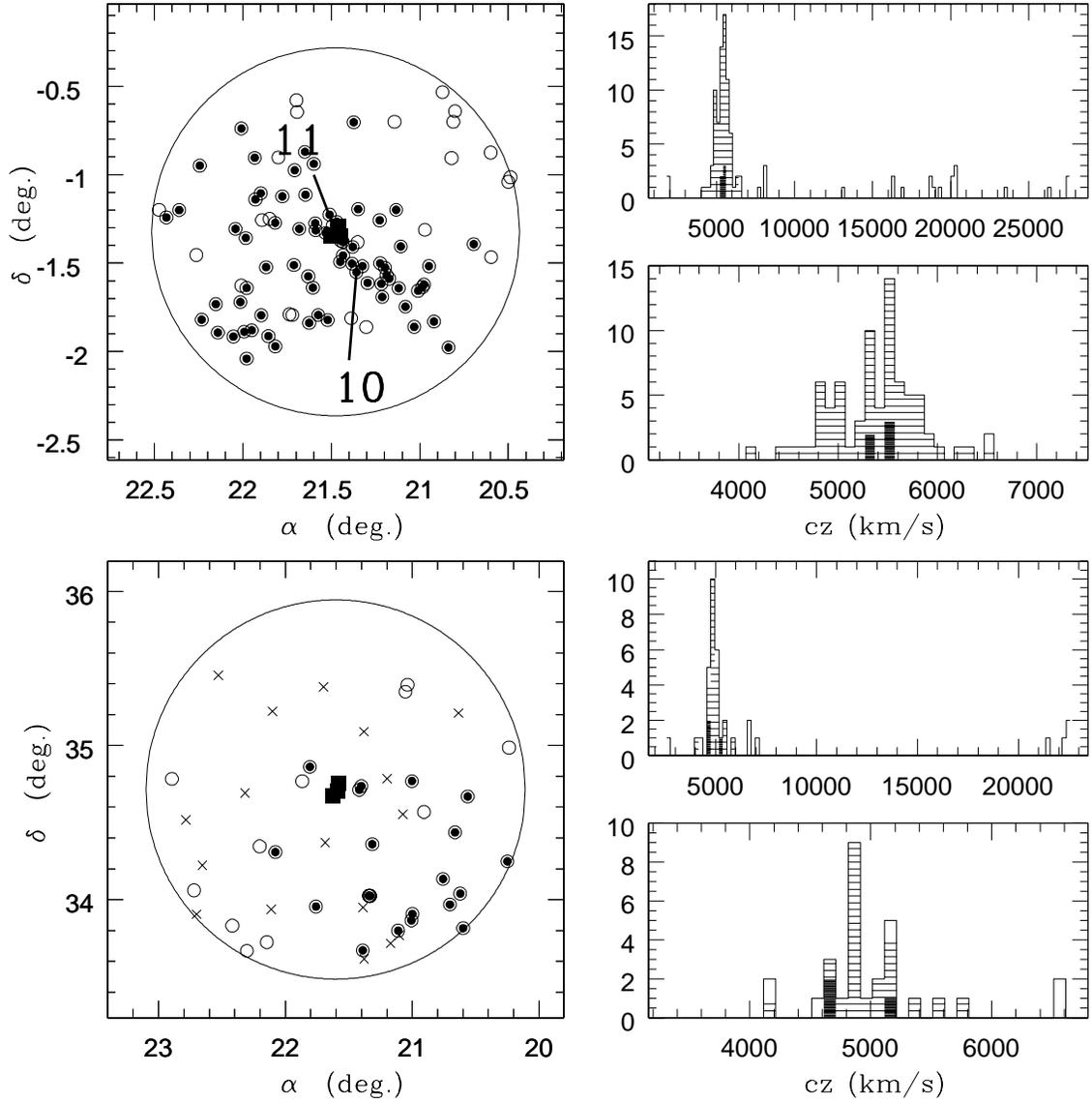}}
\caption{(a) RSCG 11 and (b) RSCG 12.
Format as in Fig.~\ref{fig:845}.}
\label{fig:919}
\end{figure}

\begin{figure}
\centerline{\epsfysize=6in%
\epsffile{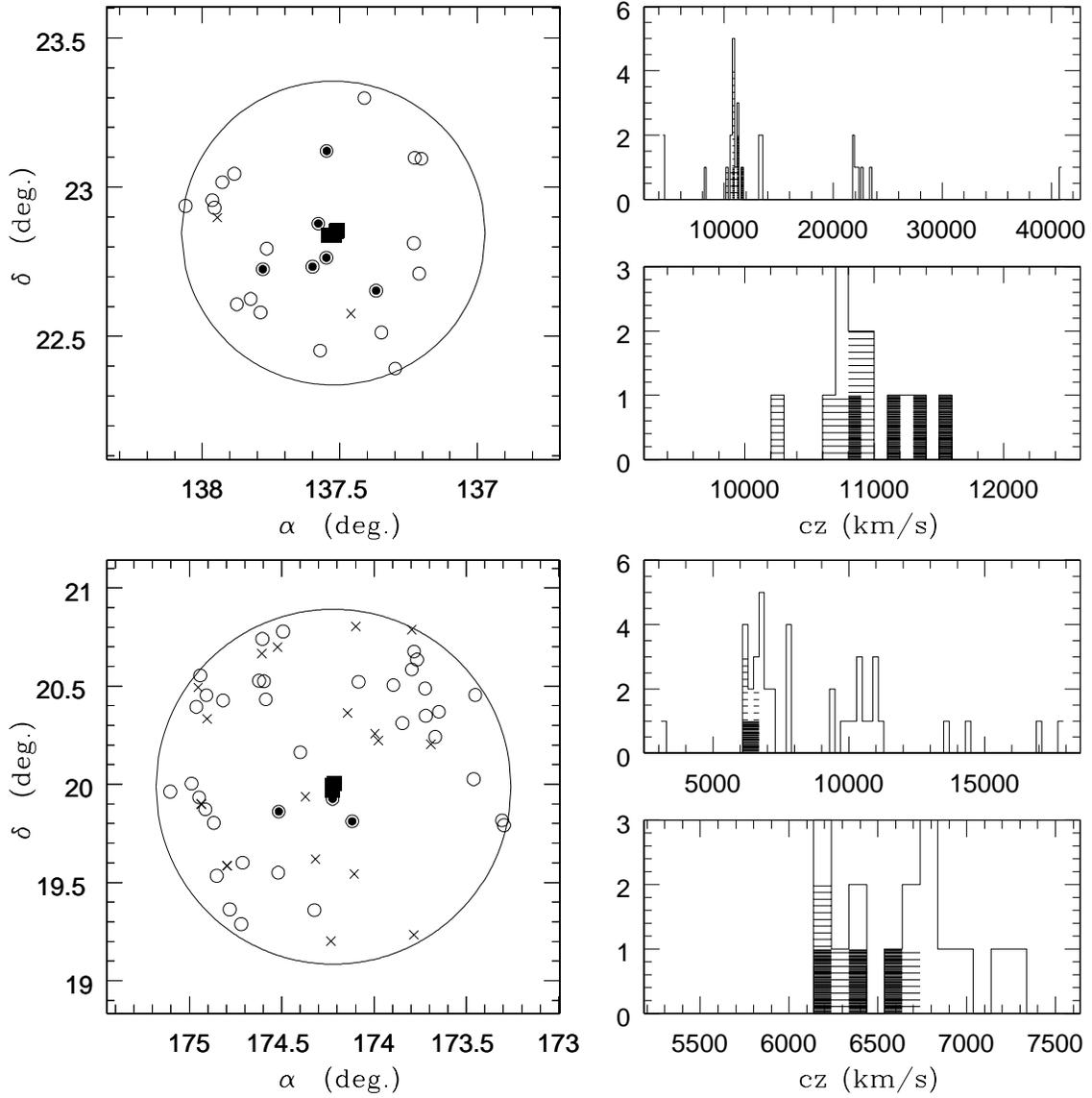}}
\caption{(a) RSCG 29 and (b) RSCG 42.
Format as in Fig.~\ref{fig:845}.}
\label{fig:5341}
\end{figure}

\begin{figure}
\centerline{\epsfysize=6in%
\epsffile{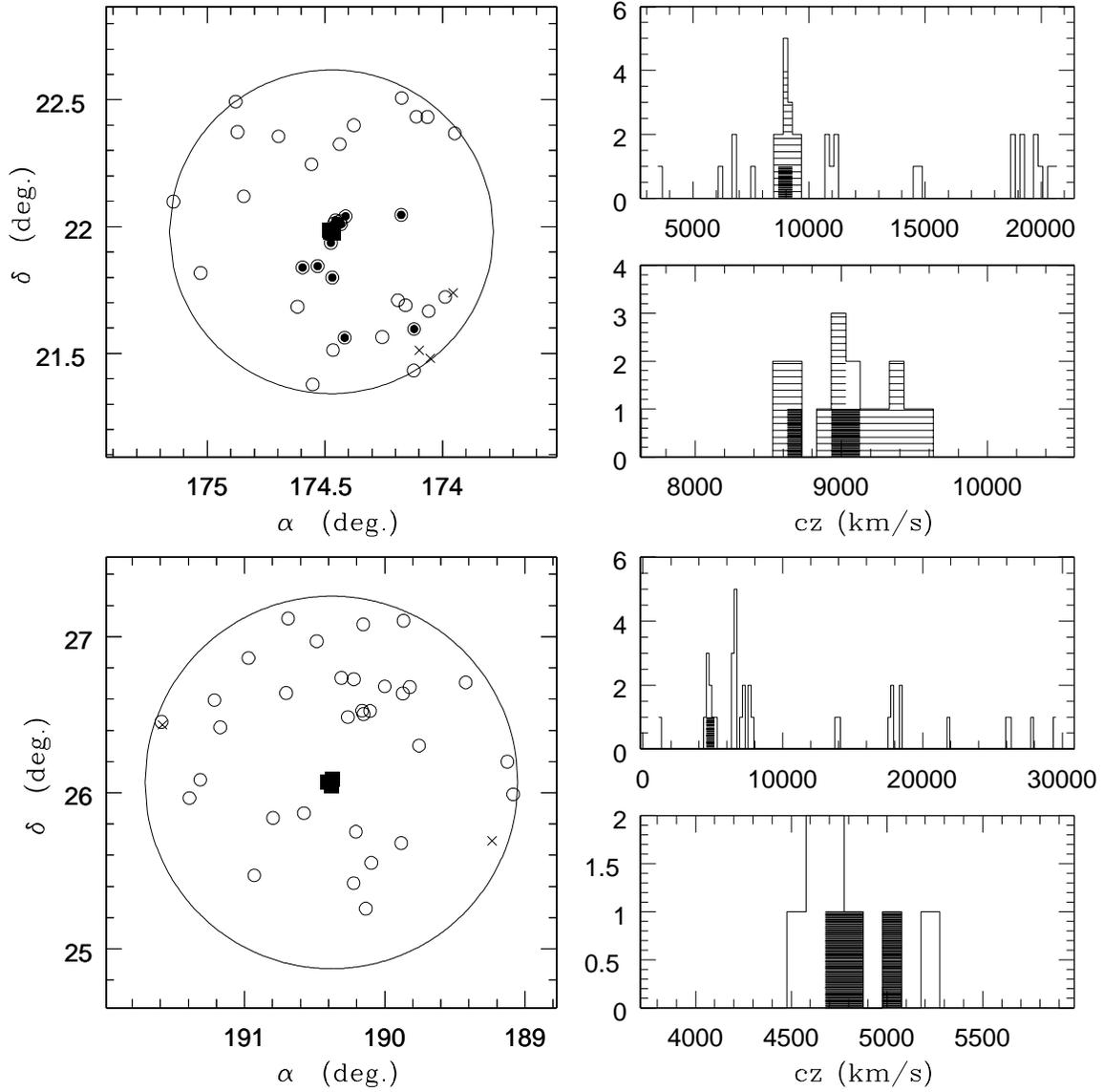}}
\caption{(a) RSCG 43 and (b) RSCG 64.
Format as in Fig.~\ref{fig:845}.}
\label{fig:4081}
\end{figure}

\begin{figure}
\centerline{\epsfysize=6in%
\epsffile{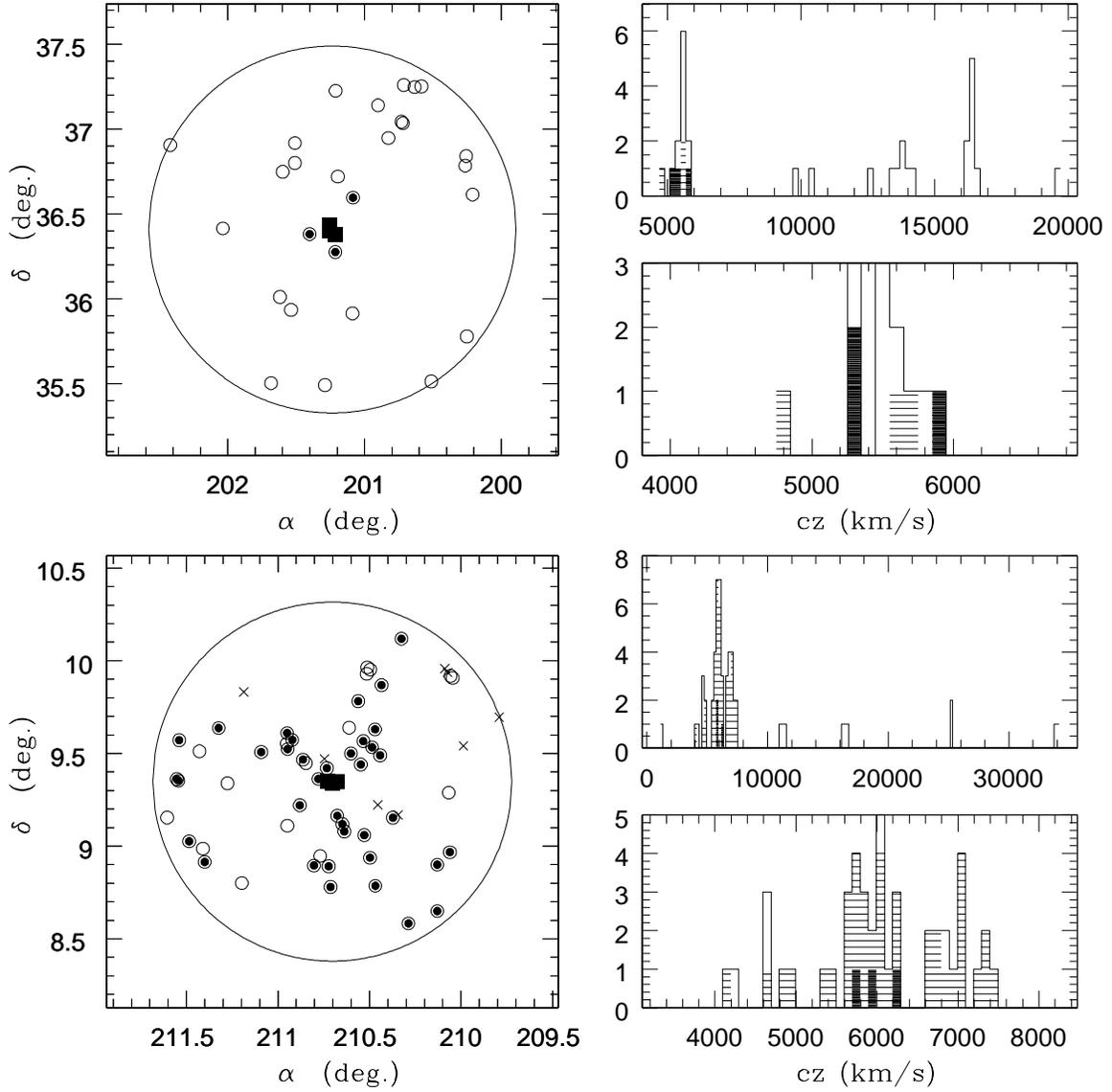}}
\caption{(a) RSCG 70 and (b) RSCG 73.
Format as in Fig.~\ref{fig:845}.}
\label{fig:1712}
\end{figure}

\begin{figure}
\centerline{\epsfysize=6in%
\epsffile{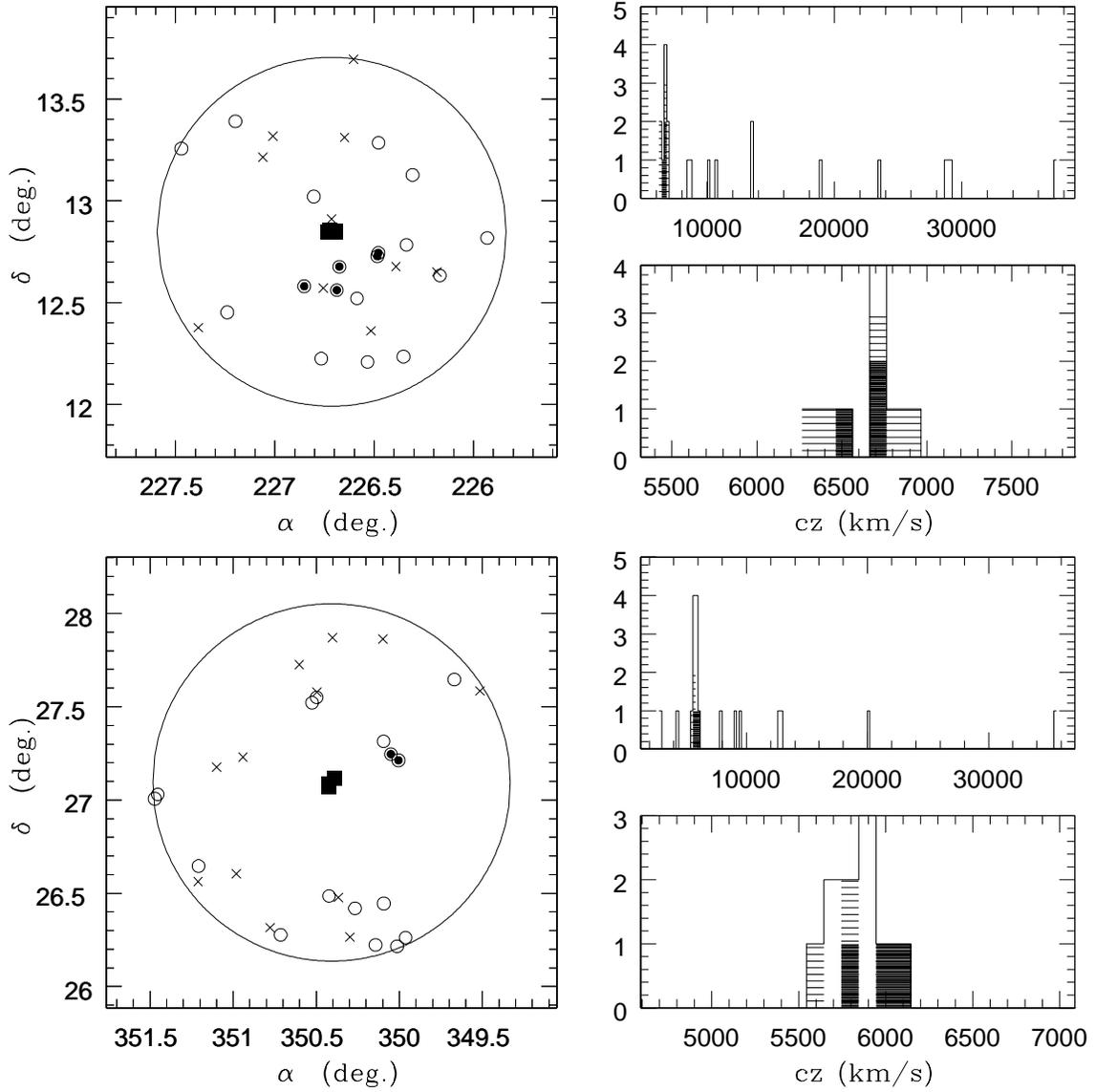}}
\caption{(a) RSCG 76 and (b) RSCG 85.
Format as in Fig.~\ref{fig:845}.}
\label{fig:76}
\end{figure}

The statistics p($\Delta v_{\rm max}$) and $D_{\rm nn,s}$ indicate that
the remaining 10 RSCGs are less distinct from their environments;
they may be bound subsystems or chance projections. Kinematic data are
inadequate to make a distinction. Next, we discuss aspects of the
individual RSCGs which we show in Figs.~\ref{fig:845}~--~\ref{fig:76}. 
 
{\it RSCG 7, RSCG 8, RSCG 12}: RSCG~7 and RSCG~8 are within the same
large, dense system of galaxies which is a very prominent feature in the
redshift survey, the Zwicky cluster (fields 501 and 502, number 5) of 625
galaxies (Zwicky \& Kowal 1968). RSCG~12, which consists of the three
tightest members of HCG~10, is on the northeast edge of this system. 

{\it RSCG~9}: RSCG~9 appears isolated on the sky and in redshift space. 
The nearest galaxy coincident in redshift space is $\sim 500 {\rm\ h}^{-1}
{\rm\ kpc}$ from the center of RSCG~9.  The velocity dispersion of RSCG~9
is the smallest in our sample ($97 \pm 49 {\rm\ km\ s}^{-1}$).  We conclude that
RSCG~9 is isolated and may be gravitationally bound. 

{\it RSCG~10, RSCG~11}: RSCG~10 and RSCG~11 are members of Abell 194, a
``linear'' cluster of galaxies (Rood \& Sastry 1971; Struble \& Rood 1982,
1984; Chapman {\it et al.} 1988).  As mentioned above, RSCG~11 is in the
core of the cluster. 

{\it RSCG~29}: RSCG~29 is the most distant RSCG in our sample ($v_{\rm
med} = 11252 {\rm\ km\ s}^{-1}$).  The Zwicky magnitudes originally listed
in the CfA redshift survey are in error and 3 of the 4 member galaxies are
actually fainter than the RSCG survey limit; the group should not have
been in our sample.  For RSCG~29, the values of $R V_0$ and $R D_0$
(Table~\ref{tab:fofa} ) are large;  the environment of the RSCG defined by
the friends-of-friends algorithm is probably overly generous.  However,
there are some close neighbors and the system appears to be embedded. 

{\it RSCG~42}: RSCG~42 is embedded in a small, loose system.  It is very
close to one of its neighbors;  because this neighbor is only 49~kpc from
one of the group members, we would have included it in the RSCG if it were
brighter.  We add this galaxy and recompute the group parameters, without
readjusting $v_{\rm med}$; the distance to the nearest neighbor is now
199~h$^{-1}$~kpc;  Table~\ref{tab:param} lists the relevant parameters
under the group heading ``RSCG~42 + 1''.  RSCG~42 + 1 is one of the more
isolated RSCGs, and may be a real compact group within a loose system. 

{\it RSCG~43}: RSCG~43 is the densest part of HCG~57, an eight-member
compact group.  Table~\ref{tab:param} lists the relevant parameters
computed with only the 3 RSCG members, with all 8 HCG members, and with an
additional nearby faint galaxy (1.7 arcmin~$\approx 45 {\rm\ h^{-1}\ kpc}$
away from an HCG member).  The three original members of the RSCG are {\it
very} tight on the sky --- the radius of the RSCG is only $\sim 13.3 {\rm\
h^{-1} \ kpc}$. 

{\it RSCG~64}:  RSCG~64 is a very tight system ($< 20 {\rm\ kpc} $ in
radius) with a low velocity dispersion ($\sigma_{\rm RSCG} = 111 \pm 74 {\rm\ km\
s}^{-1}$).  The system is near the edge of a small apparent void.  Only 5
galaxies within the entire region are roughly coincident with the RSCG in
velocity space, and the nearest of these is $560 {\rm\ h^{-1}\ kpc}$ away
from the RSCG center.  RSCG~64 is probably an isolated, gravitationally
bound system. No signs of tidal interaction are evident. 

{\it RSCG~73}: RSCG~73 is embedded in a dense system of galaxies.  The
friends-of-friends algorithm identifies 40 galaxies in its environment. 
The velocity histogram indicates that these galaxies are a superposition
of at least 2 systems along with a small number of foreground galaxies. 
In any case, RSCG~73 is not isolated. 

The range of RSCG embeddings (local environments) is qualitatively similar
in its extremes to the range of HCG embeddings. de Carvalho {\it et al.} 
(1994) searched
automated scans of IIIa-J plates in a $\frac{1}{2}^{\circ} \times
\frac{1}{2}^{\circ}$ region, to ${\rm m}_B \leq 19.5$.  They used
Hickson's (1982) compactness criterion, omitting the isolation criterion,
to redefine the compact groups, including the faint galaxies. They used
available redshifts and assigned classifications to the group
environments.  They also find a range of systems, including systems like
HCG~4 which appears relatively compact and isolated like RSCG~64, and
systems in like HCG~21 which they find in a rich environment on the
sky. In our study, RSCG~10 and RSCG~43 are both parts of HCGs (10 and 57,
respectively); here we find that HCG~10 is located on the edge of a rich
Zwicky cluster (Zwicky \& Kowal 1968).

\section{Conclusion}

We extend the CfA2 redshift survey to limiting magnitudes of $m_{\rm Zw}
\sim 16$~--~$17$ by measuring fainter galaxies within $1 {\rm h}^{-1}{\rm\
Mpc}$ of 14 RSCGs to understand the distinction between RSCGs and their
environments, and to explore the nature of the surroundings of apparent
compact groups. We define the environments of the RSCGs using the
friends-of-friends algorithm and find: 

\begin{itemize}

\item RSCGs are distinct from their environments to varying degrees;
qualitatively, the range of RSCG embeddings is similar to the range of HCG
embeddings (de Carvalho {\it et al.} 1994). 

\item One of the RSCGs is not located in an overdense region in
redshift space (RSCG~64). Of the remaining 13 RSCGs,
which are embedded in systems, 3 appear distinct from their environments in
redshift or position on the sky (RSCG~9, RSCG~11 and RSCG~85).  

\item 13 of 14 RSCGs are embedded in systems that qualify as systems that
are overdense in redshift space by the standards of Ramella {\it et al.} 
(1989).  These systems vary from a loose group of 5 members to an 
Abell cluster.

\end{itemize}

Maps of the environments of compact groups in position and redshift
provide only one limited measure of whether they are physical systems. 
These studies provide insufficient constraints on the true spatial
distribution of galaxies within loose groups or denser systems to form the
basis for extensive modeling.  Other techniques for determining
whether a compact group is a physical system are deep optical (B-band)
imaging to look for evidence of tidal interactions among group members,
studies of internal galaxy dynamics to look for distortion, spectroscopic
classification to look for star formation and nuclear activity, and x-ray
imaging to look for hot gas in the group centers.  Other 
investigators have studied
HCGs using all of these techniques (e.g. optical: Hickson {\it et al.}
1989; dynamics: Rubin {\it et al. } 1991; spectroscopic: Coziol {\it et
al.} 1997; x-ray: Ebeling {\it et al.} 1994).  Some similar studies of
RSCGs are in progress (optical: Barton {\it et al.} 1998; x-ray: Mahdavi
{\it et al.} 1998). 

\acknowledgements

We thank Susan Tokarz for reducing the spectroscopic data.  We also thank
S. G. Djorgovski for allowing us to use the POSS-II data in advance of
publication.  We thank Perry Berlind and Jim Peters for observations, and
Michael Kurtz, Emilio Falco and Massimo Ramella for their advice and
assistance.  We thank an anonymous referee for suggestions which led
us to clarify the limitations of redshift space analysis of compact 
groups. This research has made use of the NASA/IPAC Extragalactic
Database (NED).  E.~B.  acknowledges support from a National Science
Foundation graduate fellowship.

\begin{table}
\dummytable \label{tab:ks}
\end{table}

\begin{table}
\dummytable \label{tab:obs}
\end{table}

\begin{table}
\dummytable \label{tab:glx}
\end{table}

\begin{table}
\dummytable \label{tab:fofa}
\end{table}

\begin{table}
\dummytable \label{tab:param}
\end{table}

\end{document}


\setcounter{page}{25}

\begin{deluxetable}{cr}
\tablenum{1}
\tablewidth{0pc}
\tablecolumns{2}
\tablecaption{K-S Tests of Subsample of RSCGs}
\tablehead{ 
\colhead{Parameter}  & \colhead{K-S Probability } }
\startdata
Redshift & 13.3 \% \nl
Velocity Dispersion & 26.1 \% \nl
$\Upsilon$ (Isolation) & 18.9 \% \nl
Density ($n$) & 5.2 \%\nl
Linear Group Radius & 16.3 \%\nl
Environment Overdensity & 47.7 \%\nl
\enddata
\end{deluxetable}

\begin{deluxetable}{crrrrr}
\tablenum{2}
\tablewidth{0pc}
\tablecolumns{6}
\tablecaption{Environment Fields}
\tablehead{
 & \multicolumn{2}{c}{Position (J2000)} &  &  & \\ \colhead{Group} & \multicolumn{1}{c}{$\alpha$} & \multicolumn{1}{c}{$\delta$} & \colhead{ $N_{\rm gal}$} & \colhead{ $N_{\rm new}$} & \colhead{Completeness} }
\startdata
RSCG~7  & $01^h 23^m 09.2^s$ & $+33^{\circ} 27' 08''$ & 87\tablenotemark{a} & 45\tablenotemark{a}& 91\% complete to $m_{\rm Zw} \sim 16.40$ \nl
RSCG~8  & 01 23 28.2 &  +33  16  05  & 76\tablenotemark{a} & 38\tablenotemark{a} & 91\% complete to $m_{\rm Zw} \sim 16.40$\nl
RSCG~9  & 01 25 18.7 &  +14  51  34  & 29& 25& 85\% complete to $m_{\rm Zw} \sim 17.09$\nl
RSCG~10 & 01 25 25.5 &  -01  31  09  & 111\tablenotemark{b} & 49\tablenotemark{b} & 100\% complete to $m_{\rm Zw} \sim 16.46$\nl
RSCG~11 & 01 25 54.2 &  -01  19  23  &  99\tablenotemark{b} & 37\tablenotemark{b}& 100\% complete to $m_{\rm Zw} \sim 16.46$\nl
RSCG~12 & 01 26 24.9 &  +34  42  53  & 35\tablenotemark{a} & 14\tablenotemark{a} & 72\% complete to $m_{\rm Zw} \sim 16.31$\nl
RSCG~29 & 09 10 05.7 &  +22  50  47  & 27& 23& 96\% complete to $m_{\rm Zw} \sim 16.87$\nl
RSCG~42 & 11 36 53.0 &  +19  59  16  & 41& 32& 100\% complete to $m_{\rm Zw} \sim 16.78$\nl
RSCG~43 & 11 37 53.3 &  +21  58  47  & 37& 26& 92\% complete to $m_{\rm Zw} \sim 16.56$\nl
RSCG~64 & 12 41 31.2 &  +26  03  59  & 35& 24& 94\% complete to $m_{\rm Zw} \sim 16.92$\nl
RSCG~70 & 13 24 56.4 &  +36  24  28  & 30& 21& 100\% complete to $m_{\rm Zw} \sim 16.75$\nl
RSCG~73 & 14 02 48.6 &  +09  20  53  & 56& 24& 97\% complete to $m_{\rm Zw} \sim 16.28$\nl
RSCG~76 & 15 06 51.4 &  +12  50  55  & 21& 13& 88\% complete to $m_{\rm Zw} \sim 16.11$\nl
RSCG~85 & 23 21 37.8 &  +27  05  40  & 19& 13& 68\% complete to $m_{\rm Zw} \sim 16.60$\nl
\tablecomments{The composition of the environment catalogs: (1) Group number,
(2) -- (3) position, (4) number of galaxies with redshifts in each field, 
(5) number of redshifts taken for this study, and 
(6) the Zwicky magnitude limit used for analysis.}
\tablenotetext{a}{The RSCG~7, RSCG~8, and RSCG~12 regions overlap.  The combined values are: $N_{\rm gal} = 102$ and $N_{\rm new} = 49$.}
\tablenotetext{b}{The RSCG~10 and RSCG~11 regions overlap.  The combined values are: $N_{\rm gal} = 112$ and $N_{\rm new} = 50$.}
\enddata
\end{deluxetable}

\begin{deluxetable}{rrrrrl}
\tablenum{3}
\tablewidth{0pc}
\tablecolumns{6}
\tablecaption{Catalog of Newly Measured Redshifts Surrounding RSCGs}
\tablehead{
\multicolumn{2}{c}{Position (J2000)} & \colhead{Velocity} & \colhead{Error} &  & \\
\colhead{$\alpha$} & \colhead{$\delta$} & \multicolumn{2}{c}{(km s$^{-1}$)} & \colhead{Name} & \colhead{Comments} }
\startdata
& & & & & \nl
\multicolumn{6}{c}{\it        RSCGs 7, 8, 12        } \nl
& & & & & \nl
\tableline
& & & & & \nl

01 18 53.1 &   33 10 23 &    5721 &   77 &              & env. of 7,  8               \nl

01 19 10.1 &   33 01 50 &    5620 &   79 & 01163+3247   & env. of 7,  8               \nl

01 19 48.5 &   32 46 04 &    4964 &   64 &              & env. of 7,  8               \nl

01 20 01.7 &   33 39 21 &    3844 &   78 &              & env. of 7,  8               \nl

01 20 37.8 &   33 33 26 &    5334 &   78 &              & env. of 7,  8               \nl

01 20 46.4 &   33 02 42 &    5110 &   22 &              & env. of 7,  8               \nl

01 20 56.9 &   34 59 11 &    4601 &   50 & 01181+3444   &                             \nl

01 20 59.9 &   34 14 55 &    5104 &   28 &              & env. of 7,  12              \nl

01 21 03.0 &   33 53 55 &    4029 &   25 & 01182+3338   & env. of 7,  8               \nl

01 21 05.7 &   33 22 43 &    5223 &   22 & 01183+3308   & env. of 7,  8               \nl

01 21 17.5 &   33 05 26 &    5109 &   26 & 01184+3250   & env. of 7,  8               \nl

01 21 18.3 &   33 09 31 &    4086 &   62 &              & env. of 7,  8               \nl

01 21 32.3 &   32 12 53 &   10433 &   78 & 01187+3158   &                             \nl

01 21 34.9 &   33 36 01 &    4548 &   64 & 01188+3321   & env. of 7,  8               \nl

01 21 37.4 &   32 36 21 &    4745 &   21 & 01188+3221   & env. of 7,  8               \nl

01 21 44.6 &   33 29 37 &    5376 &   56 &              & env. of 7,  8               \nl

01 21 55.6 &   33 07 51 &    5366 &   77 &              & env. of 7,  8               \nl

01 22 01.0 &   33 30 35 &    5184 &   27 &              & env. of 7,  8               \nl

01 22 11.8 &   33 26 57 &    4354 &   27 &              & env. of 7,  8               \nl

01 22 15.1 &   34 40 09 &    5100 &   78 & 01194+3425   & env. of 7,  12              \nl

01 22 23.5 &   33 48 55 &    5129 &   37 &              & env. of 7,  8,  12          \nl

01 22 24.0 &   32 16 07 &   17744 &   32 &              &                             \nl

01 22 25.7 &   32 41 51 &    5333 &   26 &              & env. of 7,  8               \nl

01 22 29.2 &   34 02 24 &    4916 &   79 &              & env. of 7,  8,  12          \nl

01 22 39.0 &   34 26 13 &    4904 &   80 &              & env. of 7,  12              \nl

01 22 48.7 &   33 58 06 &    5560 &   78 &              & env. of 7,  8,  12          \nl

01 22 53.2 &   33 24 49 &    3952 &   58 &              & env. of 7,  8               \nl

01 23 06.6 &   33 11 22 &    4462 &   37 &              & env. of 7,  8               \nl

01 23 11.6 &   33 31 45 &    6030 &   78 &              & env. of 7,  8               \nl

01 23 14.8 &   33 33 46 &    4424 &   15 &              & env. of 7,  8               \nl

01 23 23.5 &   33 22 55 &    4880 &   32 &              & env. of 7,  8               \nl

01 23 27.9 &   33 12 16 &    4226 &   24 &              & in RSCG 8                   \nl

01 23 28.1 &   33 04 59 &    5023 &   27 &              & env. of 7,  8               \nl

01 23 37.5 &   32 37 49 &    4823 &   78 & 01208+3223   & env. of 7,  8               \nl

01 23 37.9 &   34 34 09 &    6993 &   78 &              &                             \nl

01 23 43.2 &   33 24 59 &    5601 &   41 &              & env. of 7,  8               \nl

01 23 50.2 &   33 35 06 &    4984 &   26 &              & env. of 7,  8               \nl

01 23 52.0 &   32 39 31 &   12140 &   43 & 01210+3225   &                             \nl

01 23 58.5 &   33 18 48 &    5031 &   21 &              & env. of 7,  8               \nl

01 24 09.0 &   35 23 35 &   22218 &   44 &              &                             \nl

01 24 12.8 &   35 20 56 &   22590 &   79 &              &                             \nl

01 24 25.8 &   33 24 25 &    4983 &   39 &              & env. of 7,  8               \nl

01 24 26.8 &   33 47 57 &    5796 &   26 &              & env. of 7,  8,  12          \nl

01 25 36.2 &   34 44 10 &    4843 &   21 &              & env. of 12                  \nl

01 25 42.6 &   33 05 17 &    4301 &   65 &              & env. of 7,  8               \nl

01 25 55.4 &   33 26 14 &   14661 &   42 &              &                             \nl

01 27 28.0 &   34 46 05 &   21494 &   81 &              &                             \nl

01 27 58.3 &   32 59 29 &    5149 &   30 & 01252+3245   & env. of 8                   \nl

01 28 35.1 &   33 43 29 &   22401 &   79 &              &                             \nl
& & & & & \nl
\tableline
& & & & & \nl
\multicolumn{6}{c}{\it            RSCG  9           } \nl
& & & & & \nl
\tableline
& & & & & \nl

01 21 53.3 &   14 53 44 &   10921 &   24 &              &                             \nl

01 23 16.9 &   14 39 32 &   11897 &   78 &              &                             \nl

01 23 34.2 &   15 04 56 &   15419 &   31 &              &                             \nl

01 23 47.6 &   14 14 33 &   21742 &   50 &              &                             \nl

01 24 02.4 &   14 25 06 &   15806 &   78 &              &                             \nl

01 24 08.2 &   14 02 04 &   16714 &   30 &              &                             \nl

01 24 08.9 &   14 02 05 &   16281 &   36 &              &                             \nl

01 24 38.5 &   15 43 44 &    2274 &   77 &              &                             \nl

01 24 54.7 &   14 13 06 &    5910 &   77 &              &                             \nl

01 25 04.6 &   14 23 59 &    5043 &   79 &              &                             \nl

01 25 07.0 &   14 53 36 &   15806 &   78 &              &                             \nl

01 25 13.3 &   14 52 21 &    6396 &   79 &              & in RSCG 9                   \nl

01 25 22.1 &   14 50 11 &    6586 &   21 &              & in RSCG 9                   \nl

01 25 24.7 &   14 51 54 &    6368 &   20 &              & in RSCG 9                   \nl

01 25 34.6 &   14 58 32 &   15765 &   83 & 01230+1443   &                             \nl

01 26 23.8 &   15 36 15 &   13377 &   80 &              &                             \nl

01 26 34.0 &   14 12 03 &   16504 &   27 &              &                             \nl

01 26 39.9 &   14 12 03 &   16785 &   79 &              &                             \nl

01 26 48.7 &   14 30 17 &   10996 &   79 &              &                             \nl

01 27 23.2 &   15 33 13 &   15697 &   79 &              &                             \nl

01 27 31.1 &   14 49 11 &    6492 &   78 & 01248+1435   & env. of 9                   \nl

01 27 38.2 &   14 21 11 &   21684 &   78 &              &                             \nl

01 28 30.9 &   14 58 52 &    7381 &   25 &              &                             \nl

01 28 42.1 &   14 36 32 &   10962 &   35 &              &                             \nl

01 28 53.8 &   14 43 14 &   11073 &   24 &              &                             \nl
& & & & & \nl
\tableline
& & & & & \nl
\multicolumn{6}{c}{\it  RSCGs 10 and 11 (Abell 194) } \nl
& & & & & \nl
\tableline
& & & & & \nl

01 21 00.8 &  -01 31 46 &   18920 &   77 &              &                             \nl

01 21 24.8 &  -01 25 21 &   28351 &   38 &              &                             \nl

01 21 37.1 &  -01 31 40 &   19048 &   77 &              &                             \nl

01 21 56.5 &  -01 59 21 &   20367 &   19 &              &                             \nl

01 22 21.4 &  -02 12 33 &   20376 &   77 &              &                             \nl

01 22 23.3 &  -01 27 58 &   18836 &   21 &              &                             \nl

01 22 27.3 &  -02 18 16 &   15974 &   42 &              &                             \nl

01 22 38.2 &  -02 22 45 &   16191 &   18 &              &                             \nl

01 23 11.3 &  -02 14 31 &   13597 &   32 &              &                             \nl

01 23 21.2 &  -01 58 37 &    4836 &   77 & 01208-0214   & env. of 10,  11             \nl

01 23 33.3 &  -02 27 22 &   13604 &   41 &              &                             \nl

01 23 47.6 &  -01 31 02 &    5211 &   21 &              & env. of 10,  11             \nl

01 23 53.3 &  -01 18 43 &   16910 &   36 &              &                             \nl

01 24 24.5 &  -02 28 58 &    5630 &   31 &              &                             \nl

01 24 32.2 &  -01 11 57 &    5050 &   19 &              & env. of 10,  11             \nl

01 24 34.4 &  -00 42 03 &   13100 &   78 &              &                             \nl

01 24 41.7 &  -01 35 16 &    5036 &   35 &              & env. of 10,  11             \nl

01 24 47.3 &  -01 31 33 &    5356 &   26 &              & env. of 10,  11             \nl

01 25 03.9 &  -02 30 07 &   12239 &   76 &              &                             \nl

01 25 10.8 &  -01 36 44 &    5285 &   76 &              & env. of 10,  11             \nl

01 25 29.8 &  -00 42 14 &    5522 &   36 &              & env. of 10,  11             \nl

01 25 40.4 &  -02 23 16 &   21638 &   50 &              &                             \nl

01 25 47.7 &  -01 20 41 &    5533 &   21 &              & in RSCG 11                  \nl

01 26 05.0 &  -01 49 18 &    5586 &   33 &              & env. of 10,  11             \nl

01 26 07.5 &  -01 19 51 &    6038 &   34 &              & env. of 10,  11             \nl

01 26 23.8 &  -00 56 20 &    5296 &   27 &              & env. of 10,  11             \nl

01 26 30.1 &  -01 50 17 &    5513 &   33 &              & env. of 10,  11             \nl

01 26 36.0 &  -00 52 15 &    6516 &   80 &              & env. of 10,  11             \nl

01 26 47.4 &  -00 34 46 &   27524 &   77 &              &                             \nl

01 26 53.1 &  -01 47 36 &   18775 &   77 &              &                             \nl

01 26 56.8 &  -01 47 21 &   18792 &   78 &              &                             \nl

01 27 06.6 &  -01 07 21 &    5330 &   32 &              & env. of 10,  11             \nl

01 27 12.1 &  -00 54 12 &   26349 &   78 &              &                             \nl

01 27 22.4 &  -02 25 45 &   24329 &   41 &              &                             \nl

01 27 24.0 &  -01 14 59 &   20136 &   80 &              &                             \nl

01 27 28.7 &  -01 31 23 &    4964 &   56 &              & env. of 10,  11             \nl

01 27 34.1 &  -01 15 23 &   20269 &   24 &              &                             \nl

01 27 35.3 &  -01 47 41 &    5814 &   34 &              & env. of 10,  11             \nl

01 27 43.0 &  -01 08 23 &    5244 &   20 & 01251-0123   & env. of 10,  11             \nl

01 27 44.0 &  -00 54 17 &    5622 &   78 &              & env. of 10,  11             \nl

01 27 55.1 &  -01 38 27 &    5663 &   29 &              & env. of 10,  11             \nl

01 27 56.1 &  -01 21 28 &    4650 &   78 &              & env. of 10,  11             \nl

01 27 58.3 &  -01 53 19 &    5536 &   79 &              & env. of 10,  11             \nl

01 28 02.1 &  -00 44 19 &    5695 &   22 &              & env. of 10,  11             \nl

01 28 02.3 &  -01 37 37 &   20046 &   65 &              &                             \nl

01 28 03.5 &  -01 43 13 &    5745 &   78 &              & env. of 10,  11             \nl

01 28 34.2 &  -01 53 34 &    4989 &   78 &              & env. of 10,  11             \nl

01 29 03.2 &  -01 27 16 &   20273 &   77 &              &                             \nl

01 29 26.6 &  -01 11 59 &    4896 &   78 &              & env. of 10,  11             \nl

01 29 54.0 &  -01 11 53 &   23479 &   79 &              &                             \nl
& & & & & \nl
\tableline
& & & & & \nl
\multicolumn{6}{c}{\it            RSCG 29           } \nl
& & & & & \nl
\tableline
& & & & & \nl

09 08 48.9 &   23 05 42 &   13206 &   78 &              &                             \nl

09 08 50.8 &   22 42 38 &    8309 &   77 &              &                             \nl

09 08 54.7 &   23 05 53 &   40901 &   38 &              &                             \nl

09 08 55.5 &   22 48 44 &    4529 &   77 &              &                             \nl

09 09 11.7 &   22 23 31 &   23544 &   79 &              &                             \nl

09 09 23.8 &   22 30 48 &    4451 &   25 &              &                             \nl

09 09 28.4 &   22 39 12 &   10963 &   34 &              & env. of 29                  \nl

09 09 38.5 &   23 17 55 &   22208 &   26 &              &                             \nl

09 10 11.4 &   23 07 17 &   10844 &   31 &              & env. of 29                  \nl

09 10 11.8 &   22 45 49 &   10219 &   78 &              & env. of 29                  \nl

09 10 17.1 &   22 27 09 &   22681 &   78 &              &                             \nl

09 10 18.8 &   22 52 41 &   10783 &   26 &              & env. of 29                  \nl

09 10 23.8 &   22 44 01 &   10910 &   79 &              & env. of 29                  \nl

09 11 03.7 &   22 47 38 &   21936 &   81 &              &                             \nl

09 11 07.2 &   22 43 31 &   10606 &   48 &              & env. of 29                  \nl

09 11 09.1 &   22 34 48 &   21821 &   40 &              &                             \nl

09 11 17.9 &   22 37 32 &   22070 &   60 &              &                             \nl

09 11 29.9 &   22 36 27 &   13365 &   80 &              &                             \nl

09 11 32.0 &   23 02 42 &   10784 &   23 &              &                             \nl

09 11 42.5 &   23 00 58 &   10736 &   42 &              &                             \nl

09 11 49.3 &   22 55 52 &   13371 &   77 &              &                             \nl

09 11 51.2 &   22 57 23 &   13298 &   77 &              &                             \nl

09 12 14.6 &   22 56 13 &   11213 &   81 &              &                             \nl
& & & & & \nl
\tableline
& & & & & \nl
\multicolumn{6}{c}{\it            RSCG 42           } \nl
& & & & & \nl
\tableline
& & & & & \nl

11 33 11.2 &   19 47 32 &    9809 &   27 &              &                             \nl

11 33 13.6 &   19 48 58 &    9950 &   80 &              &                             \nl

11 33 48.4 &   20 27 18 &   10390 &   34 &              &                             \nl

11 33 50.5 &   20 01 36 &   10570 &   28 &              &                             \nl

11 34 35.6 &   20 22 10 &   10806 &   41 & 11318+2038   &                             \nl

11 34 40.4 &   20 14 29 &    9432 &   38 &              &                             \nl

11 34 52.8 &   20 20 57 &   10464 &   30 &              &                             \nl

11 34 53.5 &   20 29 16 &    6836 &   78 & 11321+2045   &                             \nl

11 35 04.3 &   20 38 05 &    7219 &   77 &              &                             \nl

11 35 08.0 &   20 40 35 &    9290 &   78 & 11324+2057   &                             \nl

11 35 11.1 &   20 35 05 &   13605 &   37 &              &                             \nl

11 36 20.1 &   20 31 17 &    6402 &   40 & 11336+2048   &                             \nl

11 36 28.4 &   19 48 41 &    6640 &   21 & 11338+2005   & env. of 42                  \nl

11 36 51.6 &   20 00 17 &    6342 &   21 & 11342+2017   & in RSCG 42                  \nl

11 36 54.0 &   19 55 34 &    6247 &   77 &              & env. of 42                  \nl

11 37 17.6 &   19 21 36 &   14370 &   78 &              &                             \nl

11 37 58.3 &   20 46 41 &    6926 &   27 & 11354+2104   &                             \nl

11 38 03.9 &   19 51 41 &    6182 &   78 &              & env. of 42                  \nl

11 38 04.6 &   19 33 01 &   17891 &   32 &              &                             \nl

11 38 20.6 &   20 25 59 &    7260 &   38 &              &                             \nl

11 38 23.0 &   20 31 29 &    7735 &   32 & 11358+2049   &                             \nl

11 38 29.5 &   20 31 39 &    7712 &   27 & 11360+2049   &                             \nl

11 38 51.0 &   19 36 04 &    6837 &   84 & 11362+1952   &                             \nl

11 38 52.7 &   19 17 17 &   11022 &   22 &              &                             \nl

11 39 16.2 &   20 25 38 &    7757 &   36 & 11367+2043   &                             \nl

11 39 24.5 &   19 32 04 &    6779 &   25 & 11368+1948   &                             \nl

11 39 28.4 &   19 48 13 &   17042 &   37 &              &                             \nl

11 39 37.9 &   20 27 14 &    6990 &   24 &              &                             \nl

11 39 46.0 &   20 33 14 &    6188 &   40 &              &                             \nl

11 39 50.9 &   20 23 41 &    6774 &   25 &              &                             \nl

11 39 57.2 &   20 00 13 &   10127 &   79 &              &                             \nl

11 40 24.9 &   19 57 47 &    6749 &   78 &              &                             \nl
& & & & & \nl
\tableline
& & & & & \nl
\multicolumn{6}{c}{\it            RSCG 43           } \nl
& & & & & \nl
\tableline
& & & & & \nl

11 35 47.5 &   22 22 04 &   18715 &   35 &              &                             \nl

11 35 57.4 &   21 43 21 &   10964 &   35 &              &                             \nl

11 36 14.1 &   21 39 58 &   10854 &   80 &              &                             \nl

11 36 15.4 &   22 25 57 &    3579 &   78 & 11336+2242   &                             \nl

11 36 26.7 &   22 25 59 &   19764 &   45 &              &                             \nl

11 36 37.7 &   21 41 23 &   18779 &   77 &              &                             \nl

11 36 41.7 &   22 30 26 &   19924 &   78 &              &                             \nl

11 36 42.2 &   22 02 46 &    9208 &   80 &              & env. of 43                  \nl

11 36 45.6 &   21 42 36 &    7587 &   81 &              &                             \nl

11 37 01.6 &   21 33 52 &   10867 &   41 &              &                             \nl

11 37 30.4 &   22 23 59 &    6876 &   80 & 11350+2241   &                             \nl

11 37 39.0 &   22 02 26 &    8697 &   25 &              & env. of 43                  \nl

11 37 39.8 &   21 33 42 &    8623 &   32 &              & env. of 43                  \nl

11 37 44.9 &   22 19 30 &   14575 &   42 &              &                             \nl

11 37 51.6 &   21 30 48 &   11207 &   88 &              &                             \nl

11 37 52.5 &   21 47 57 &    8629 &   28 & 11354+2205   & env. of 43                  \nl

11 38 07.5 &   21 50 31 &    9522 &   21 &              & env. of 43                  \nl

11 38 12.4 &   21 22 39 &   19838 &   80 &              &                             \nl

11 38 13.6 &   22 14 44 &   14777 &   32 &              &                             \nl

11 38 22.8 &   21 50 21 &    9343 &   41 &              & env. of 43                  \nl

11 38 27.9 &   21 41 03 &   19117 &   25 &              &                             \nl

11 38 47.3 &   22 21 19 &   20585 &   79 &              &                             \nl

11 39 29.2 &   22 22 23 &    9053 &   79 &              &                             \nl

11 39 31.1 &   22 29 33 &    6848 &   78 &              &                             \nl

11 40 06.8 &   21 49 02 &   19252 &   79 &              &                             \nl

11 40 34.5 &   22 05 55 &   20355 &   78 &              &                             \nl
& & & & & \nl
\tableline
& & & & & \nl
\multicolumn{6}{c}{\it            RSCG 64           } \nl
& & & & & \nl
\tableline
& & & & & \nl

12 39 01.2 &   26 18 10 &    7822 &   79 &              &                             \nl

12 39 29.1 &   26 38 12 &   27752 &   37 &              &                             \nl

12 39 31.7 &   25 40 39 &   13956 &   46 &              &                             \nl

12 40 00.0 &   26 40 56 &    4756 &   82 &              &                             \nl

12 40 23.0 &   25 33 04 &    6392 &   54 &              &                             \nl

12 40 25.2 &   26 31 27 &    7685 &   20 &              &                             \nl

12 40 32.4 &   25 15 29 &   18524 &   27 &              &                             \nl

12 40 36.3 &   26 30 16 &    6618 &   79 &              &                             \nl

12 40 36.9 &   27 04 44 &    6661 &   78 &              &                             \nl

12 40 38.6 &   26 31 33 &    6485 &   77 &              &                             \nl

12 40 49.4 &   25 45 01 &   17866 &   27 &              &                             \nl

12 40 52.9 &   26 43 40 &    4644 &   70 &              &                             \nl

12 40 53.0 &   25 25 19 &   21927 &   40 &              &                             \nl

12 41 03.1 &   26 29 06 &   18449 &   78 &              &                             \nl

12 41 37.3 &   26 04 17 &    4711 &   78 &              & in RSCG 64                  \nl

12 41 56.2 &   26 58 16 &    7206 &   77 &              &                             \nl

12 42 18.2 &   25 52 09 &   29521 &   78 &              &                             \nl

12 43 11.1 &   25 50 20 &   17599 &   37 &              &                             \nl

12 43 43.2 &   25 28 18 &    5259 &   81 &              &                             \nl

12 44 41.4 &   26 25 12 &    4627 &   78 & 12421+2642   &                             \nl

12 44 51.5 &   26 35 39 &   13735 &   36 &              &                             \nl

12 45 15.7 &   26 05 02 &   17791 &   43 &              &                             \nl

12 45 34.2 &   25 58 00 &   26026 &   79 &              &                             \nl

12 46 21.9 &   26 27 16 &   26152 &   42 &              &                             \nl
& & & & & \nl
\tableline
& & & & & \nl
\multicolumn{6}{c}{\it            RSCG 70           } \nl
& & & & & \nl
\tableline
& & & & & \nl

13 20 50.4 &   36 36 49 &    9778 &   40 & 13186+3652   &                             \nl

13 21 01.7 &   36 50 25 &   16337 &   78 &              &                             \nl

13 22 20.3 &   37 15 03 &   13873 &   25 &              &                             \nl

13 22 32.4 &   37 14 47 &   16515 &   77 &              &                             \nl

13 22 51.1 &   37 15 32 &   16345 &   28 &              &                             \nl

13 22 52.7 &   37 02 02 &   13423 &   77 &              &                             \nl

13 22 55.3 &   37 02 32 &   13502 &   27 &              &                             \nl

13 23 18.1 &   36 56 49 &   10306 &   77 &              &                             \nl

13 23 36.2 &   37 08 21 &   13959 &   55 & 13214+3724   &                             \nl

13 24 46.9 &   36 43 08 &   16147 &   47 &              &                             \nl

13 24 50.7 &   37 13 27 &   16241 &   78 &              &                             \nl

13 24 51.3 &   36 16 31 &    5574 &   78 &              & env. of 70                  \nl

13 25 01.3 &   36 26 13 &    5877 &   77 &              & in RSCG 70                  \nl

13 25 09.3 &   35 29 28 &   12633 &   79 &              &                             \nl

13 25 36.4 &   36 22 51 &    5673 &   80 &              & env. of 70                  \nl

13 26 02.1 &   36 54 58 &   16428 &   78 &              &                             \nl

13 26 02.4 &   36 47 58 &   16332 &   24 &              &                             \nl

13 26 23.7 &   36 44 51 &   14238 &   25 &              &                             \nl

13 26 44.0 &   35 30 10 &   19669 &   44 &              &                             \nl

13 28 08.7 &   36 24 51 &   13733 &   27 &              &                             \nl

13 29 41.0 &   36 54 19 &    5509 &   31 &              &                             \nl
& & & & & \nl
\tableline
& & & & & \nl
\multicolumn{6}{c}{\it            RSCG 73           } \nl
& & & & & \nl
\tableline
& & & & & \nl

14 00 11.0 &   09 54 26 &   16543 &   79 &              &                             \nl

14 00 13.8 &   09 55 05 &   16742 &   32 &              &                             \nl

14 01 08.9 &   08 35 01 &    4876 &   26 & 13587+0851   & env. of 73                  \nl

14 01 29.3 &   09 09 13 &    4927 &   78 &              & env. of 73                  \nl

14 01 44.2 &   09 52 06 &    6613 &   22 & 13592+1007   & env. of 73                  \nl

14 01 56.7 &   09 32 00 &    5644 &   24 &              & env. of 73                  \nl

14 01 59.1 &   09 57 09 &   11140 &   26 &              &                             \nl

14 02 02.9 &   09 57 51 &   11227 &   83 &              &                             \nl

14 02 06.8 &   09 03 35 &    5765 &   78 &              & env. of 73                  \nl

14 02 08.2 &   09 33 51 &    5709 &   77 &              & env. of 73                  \nl

14 02 14.8 &   09 46 56 &    6760 &   21 &              & env. of 73                  \nl

14 02 24.4 &   09 29 59 &    6122 &   35 &              & env. of 73                  \nl

14 02 26.8 &   09 38 21 &   25315 &   31 &              &                             \nl

14 02 42.1 &   09 20 47 &    6241 &   19 &              & in RSCG 73                  \nl

14 02 48.6 &   09 20 28 &    5917 &   22 &              & in RSCG 73                  \nl

14 02 55.1 &   09 20 59 &    5750 &   22 &              & in RSCG 73                  \nl

14 03 47.7 &   09 06 37 &   33943 &   40 &              &                             \nl

14 03 48.3 &   09 33 12 &   25234 &   34 &              &                             \nl

14 04 22.0 &   09 30 29 &    7314 &   78 &              & env. of 73                  \nl

14 04 47.4 &   08 48 03 &    1244 &   78 & 14023+0904   &                             \nl

14 05 38.5 &   08 59 10 &   34064 &   47 &              &                             \nl

14 06 09.5 &   09 34 18 &    7476 &   77 &              & env. of 73                  \nl

14 06 12.5 &   09 21 41 &    7378 &   29 &              & env. of 73                  \nl

14 06 25.2 &   09 09 14 &   11454 &   80 &              &                             \nl
& & & & & \nl
\tableline
& & & & & \nl
\multicolumn{6}{c}{\it            RSCG 76           } \nl
& & & & & \nl
\tableline
& & & & & \nl

15 03 43.4 &   +1 49 03 &   23549 &   78 &              &                             \nl

15 05 13.6 &   +1 07 38 &   37319 &   38 &              &                             \nl

15 05 20.9 &   +1 47 03 &    8676 &   50 &              &                             \nl

15 05 24.5 &   +1 14 05 &   13585 &   38 &              &                             \nl

15 05 54.7 &   +1 17 07 &   13590 &   78 &              &                             \nl

15 06 08.1 &   +1 12 29 &   10741 &   77 &              &                             \nl

15 06 20.9 &   +1 31 15 &   10079 &   18 &              &                             \nl

15 06 42.0 &   12 40 35 &    6692 &   79 & 15044+1253   & env. of 76                  \nl

15 06 45.2 &   12 33 41 &    6872 &   78 & 15044+1246   & env. of 76                  \nl

15 07 04.0 &   +1 13 21 &   28933 &   83 &              &                             \nl

15 07 12.9 &   +1 01 15 &   29134 &   51 &              &                             \nl

15 08 47.6 &   +1 23 25 &   18880 &   78 &              &                             \nl

15 08 57.6 &   +1 27 09 &   28788 &   45 &              &                             \nl
& & & & & \nl
\tableline
& & & & & \nl
\multicolumn{6}{c}{\it            RSCG 85           } \nl
& & & & & \nl
\tableline
& & & & & \nl

23 18 40.5 &   27 38 45 &   20008 &   82 &              &                             \nl

23 20 01.3 &   27 12 47 &    5592 &   26 &              & env. of 85                  \nl

23 20 12.3 &   27 14 45 &    5754 &   26 &              & env. of 85                  \nl

23 20 22.4 &   26 26 43 &    9416 &   26 &              &                             \nl

23 20 34.4 &   26 13 25 &    5859 &   30 &              &                             \nl

23 21 03.9 &   26 25 10 &    5859 &   77 & 23186+2608   &                             \nl

23 21 33.5 &   27 07 05 &    5981 &   78 & 23191+2651   & in RSCG 85                  \nl

23 21 41.3 &   27 05 14 &    6087 &   78 & 23192+2649   & in RSCG 85                  \nl

23 21 41.5 &   26 29 09 &    9165 &   80 &              &                             \nl

23 21 42.1 &   27 04 14 &    5782 &   80 & 23192+2648   & in RSCG 85                  \nl

23 22 51.4 &   26 16 36 &   12621 &   81 &              &                             \nl

23 24 50.0 &   26 38 45 &   12833 &   36 & 23223+2622   &                             \nl

23 25 53.2 &   27 00 26 &    5691 &   34 &              &                             \nl
& & & & & \nl
\tableline
\tablecomments{Galaxies around RSCGs: (1) -- (2) Position, (3) heliocentric velocity, (4) error, (5) name in ZCAT, and (6) comments.}
\enddata
\end{deluxetable}

\begin{deluxetable}{crrr}
\tablenum{4}
\tablewidth{0pc}
\tablecolumns{4}
\tablecaption{}
\tablehead{
 & & $R D_0$ & $R V_0$ \\
Group & $m_{\rm lim, Zw}$ & (kpc) & (km s$^{-1}$) }
\startdata
RSCG~7  & 16.40  & 474 & 468 \nl 
RSCG~8  & 16.40  & 506 & 500 \nl 
RSCG~9  & 17.09  & 484 & 495 \nl 
RSCG~10 & 16.46  & 496 & 492 \nl
RSCG~11 & 16.46  & 511 & 507 \nl
RSCG~12 & 16.31  & 492 & 484 \nl
RSCG~29 & 16.87  & 555 & 849 \nl
RSCG~42 & 16.78  & 391 & 591 \nl
RSCG~43 & 16.56  & 517 & 772 \nl
RSCG~64 & 16.92  & 332 & 509 \nl
RSCG~70 & 16.75  & 361 & 544 \nl
RSCG~73 & 16.28  & 422 & 617 \nl
RSCG~76 & 16.11  & 483 & 696 \nl
RSCG~85 & 16.60  & 515 & 514 \nl
\tablecomments{Parameters used to determine the Zwicky magnitude limit
of the region and define the environment of the RSCG: (1) Group, (2) the 
faintest Zwicky magnitude of a galaxy with a redshift
in the catalog, (3) the resulting value of $R D_0$ used to deterimine the
environment of the RSCG, and (4) the resulting value of $R V_0$ used
to determine the environment of the RSCG.}
\enddata
\end{deluxetable}

\begin{deluxetable}{crrrrrrrrr}
\tablenum{5}
\tablewidth{0pc}
\tablecolumns{10}
\tablecaption{Environment Characteristics}
\tablehead{
 & & & \colhead{ $v_{\rm med}$} & \colhead{ $v_{\rm med,env}$} & \colhead{ $\sigma_{\rm RSCG}$} &  & \colhead{$r_{RSCG}$} & \colhead{$D_{\rm nn}$} & \colhead{$D_{\rm nn,s}$} \\
Group & $N_{\rm gal}$ & $N_{\rm env}$ & \colhead{ (km/s)} & \colhead{ (km/s)} & \colhead{ (km/s)} & \colhead{ p($\Delta v_{\rm max}$) } &\colhead{ (kpc)} & \colhead{(kpc)} & \colhead{ (scaled kpc)}}
\startdata
RSCG  7     & 3 & 73 &   4375 &   4983 &    321 $\pm$     90 & 0.428 &  38.1 &  51.8 &  38.7 \nl
RSCG  8     & 4 & 71 &   5217 &   4984 &    658 $\pm$     31 & 0.896 &  58.0 &  30.0 &  21.0 \nl
RSCG  9     & 3 &  5 &   6396 &   6441 &     97 $\pm$     49 & 1.000 &  28.1 & 507.8 & 358.9 \nl
RSCG 10     & 3 & 75 &   5472 &   5417 &    381 $\pm$     32 & 0.720 &  30.3 &  93.5 &  66.5 \nl
RSCG 11     & 5 & 75 &   5509 &   5419 &    100 $\pm$     24 & 0.004 &  33.4 &  53.6 &  37.1 \nl
RSCG 12     & 3 & 24 &   4660 &   4880 &    266 $\pm$     44 & 0.616 &  36.0 & 124.1 &  89.8 \nl
RSCG 29     & 4 & 10 &  11252 &  10877 &    275 $\pm$     33 & 0.524 &  32.6 & 117.0 &  48.3 \nl
RSCG 42     & 3 &  6 &   6342 &   6294 &    166 $\pm$     32 & 0.500 &  19.8 &  68.5 &  40.6 \nl
RSCG 42 + 1 & 4 &  6 &   6294 &   6294 &    159 $\pm$     39 & 0.200 &  43.6 & 199.2 & 117.8 \nl
RSCG 43     & 3 & 15 &   8977 &   9022 &    149 $\pm$     44 & 0.218 &  13.3 &  54.1 &  24.5 \nl
RSCG 43 + 5 & 8 & 15 &   9052 &   9022 &    257 $\pm$     34 & 0.303 &  73.0 & 116.4 &  52.8 \nl
RSCG 43 + 6 & 9 & 15 &   9022 &   9022 &    278 $\pm$     28 & 0.341 &  93.7 & 260.4 & 118.0 \nl
RSCG 64     & 3 &  3 &   4801 &   4801 &    111 $\pm$     74 & 1.000 &  19.7 & N/A   &  N/A  \nl
RSCG 70     & 3 &  6 &   5303 &   5438 &    277 $\pm$     76 & 0.550 &  31.0 & 123.5 &  79.4 \nl
RSCG 73     & 3 & 40 &   5917 &   6036 &    204 $\pm$     24 & 0.125 &  27.5 &  80.0 &  45.3 \nl 
RSCG 76     & 3 &  8 &   6688 &   6690 &    105 $\pm$     32 & 0.250 &  23.8 & 205.7 & 103.4 \nl
RSCG 85     & 3 &  5 &   5981 &   5782 &    126 $\pm$     91 & 0.300 &  29.9 & 366.9 & 249.9 \nl
\tablecomments{Parameters of RSCG environments: (1) Group,
(2) number of galaxies in RSCG (3) number of galaxies in the larger system 
within 1~h$^{-1}$~Mpc of the RSCG center, (4) median heliocentric 
velocity of the RSCG, 
(5) median heliocentric velocity of the environment, 
(6) velocity dispersion of the RSCG, (7) p($\Delta v_{\rm max}$), (8) linear 
projected radius of the
RSCG, (9) distance to the nearest neighbor, and (10) distance to the nearest neighbor, scaled by $R$.}
\enddata
\end{deluxetable}